\message
{JNL.TEX version 0.92 as of 6/9/87.  Report bugs and problems to Doug Eardley.}

\catcode`@=11
\expandafter\ifx\csname inp@t\endcsname\relax\let\inp@t=\input
\def\input#1 {\expandafter\ifx\csname #1IsLoaded\endcsname\relax
\inp@t#1%
\expandafter\def\csname #1IsLoaded\endcsname{(#1 was previously loaded)}
\else\message{\csname #1IsLoaded\endcsname}\fi}\fi
\catcode`@=12







\def\beginlinemode{\endmode
  \begingroup\parskip=0pt \obeylines\def\\{\par}\def\endmode{\par\endgroup}}
\def\beginparmode{\endmode
  \begingroup \def\endmode{\par\endgroup}}
\let\endmode=\par
{\obeylines\gdef\
{}}
\def\singlespace{\baselineskip=\normalbaselineskip}

\def\oneandahalfspace{\baselineskip=\normalbaselineskip
  \multiply\baselineskip by 3 \divide\baselineskip by 2}
\def\doublespace{\baselineskip=\normalbaselineskip \multiply\baselineskip by 2}






\def\title			
  {\null\vskip 3pt plus 0.2fill
   \beginlinemode \doublespace \raggedcenter \bf}

\def\author			
  {\vskip 3pt plus 0.2fill \beginlinemode
   \singlespace \raggedcenter\sc}

\def\affil			
  {\vskip 3pt plus 0.1fill \beginlinemode
   \oneandahalfspace \raggedcenter \sl}

\def\abstract			
  {\vskip 3pt plus 0.3fill \beginparmode
   \oneandahalfspace ABSTRACT: }

\def\endtitlepage		
  {\endpage			
   \body}

\def\body			
  {\beginparmode}		

\def\subhead#1{			
  \vskip 0.25truein		
  {\raggedcenter {#1} \par}
   \nobreak\vskip 0.25truein\nobreak}

\def\beginitems{
\par\medskip\bgroup\def\i##1 {\item{##1}}\def\ii##1 {\itemitem{##1}}
\leftskip=36pt\parskip=0pt}
\def\enditems{\par\egroup}

\def\beneathrel#1\under#2{\mathrel{\mathop{#2}\limits_{#1}}}

\def\refto#1{~[{#1}]}

\def\references			
  {
   \beginparmode
   \frenchspacing \parindent=0pt \leftskip=1truecm
   \parskip=8pt plus 3pt \everypar{\hangindent=\parindent}}

\gdef\refis#1{\item{#1.\ }}			

\gdef\journal#1, #2, #3, 1#4#5#6{		
    {\sl #1~}{\bf #2}, #3 (1#4#5#6)}		

\def\endreferences{\body}

\catcode`@=11
\newcount\r@fcount \r@fcount=0
\newcount\r@fcurr
\immediate\newwrite\reffile
\newif\ifr@ffile\r@ffilefalse
\def\w@rnwrite#1{\ifr@ffile\immediate\write\reffile{#1}\fi\message{#1}}

\def\writer@f#1>>{}
\def\referencefile{
  \r@ffiletrue\immediate\openout\reffile=\jobname.ref%
  \def\writer@f##1>>{\ifr@ffile\immediate\write\reffile%
    {\noexpand\refis{##1} = \csname r@fnum##1\endcsname = %
     \expandafter\expandafter\expandafter\strip@t\expandafter%
     \meaning\csname r@ftext\csname r@fnum##1\endcsname\endcsname}\fi}%
  \def\strip@t##1>>{}}

\def\citeall#1{\xdef#1##1{#1{\noexpand\cite{##1}}}}
\def\cite#1{\each@rg\citer@nge{#1}}	

\def\each@rg#1#2{{\let\thecsname=#1\expandafter\first@rg#2,\end,}}
\def\first@rg#1,{\thecsname{#1}\apply@rg}	
\def\apply@rg#1,{\ifx\end#1\let\next=\relax
\else,\thecsname{#1}\let\next=\apply@rg\fi\next}

\def\citer@nge#1{\citedor@nge#1-\end-}	
\def\citer@ngeat#1\end-{#1}
\def\citedor@nge#1-#2-{\ifx\end#2\r@featspace#1 
  \else\citel@@p{#1}{#2}\citer@ngeat\fi}	
\def\citel@@p#1#2{\ifnum#1>#2{\errmessage{Reference range #1-#2\space is bad.}%
    \errhelp{If you cite a series of references by the notation M-N, then M and
    N must be integers, and N must be greater than or equal to M.}}\else%
 {\count0=#1\count1=#2\advance\count1 by1\relax\expandafter\r@fcite\the\count0,%
  \loop\advance\count0 by1\relax
    \ifnum\count0<\count1,\expandafter\r@fcite\the\count0,%
  \repeat}\fi}

\def\r@featspace#1#2 {\r@fcite#1#2,}	
\def\r@fcite#1,{\ifuncit@d{#1}
    \newr@f{#1}%
    \expandafter\gdef\csname r@ftext\number\r@fcount\endcsname%
                     {\message{Reference #1 to be supplied.}%
                      \writer@f#1>>#1 to be supplied.\par}%
 \fi%
 \csname r@fnum#1\endcsname}
\def\ifuncit@d#1{\expandafter\ifx\csname r@fnum#1\endcsname\relax}%
\def\newr@f#1{\global\advance\r@fcount by1%
    \expandafter\xdef\csname r@fnum#1\endcsname{\number\r@fcount}}

\let\r@fis=\refis			
\def\refis#1#2#3\par{\ifuncit@d{#1}
   \newr@f{#1}%
   \w@rnwrite{Reference #1=\number\r@fcount\space is not cited up to now.}\fi%
  \expandafter\gdef\csname r@ftext\csname r@fnum#1\endcsname\endcsname%
  {\writer@f#1>>#2#3\par}}

\def\ignoreuncited{
   \def\refis##1##2##3\par{\ifuncit@d{##1}%
     \else\expandafter\gdef\csname r@ftext\csname r@fnum##1\endcsname\endcsname%
     {\writer@f##1>>##2##3\par}\fi}}

\def\r@ferr{\endreferences\errmessage{I was expecting to see
\noexpand\endreferences before now;  I have inserted it here.}}
\let\r@ferences=\references
\def\references{\r@ferences\def\endmode{\r@ferr\par\endgroup}}

\let\endr@ferences=\endreferences
\def\endreferences{\r@fcurr=0
  {\loop\ifnum\r@fcurr<\r@fcount
    \advance\r@fcurr by 1\relax\expandafter\r@fis\expandafter{\number\r@fcurr}%
    \csname r@ftext\number\r@fcurr\endcsname%
  \repeat}\gdef\r@ferr{}\endr@ferences}


\let\r@fend=\endpaper\gdef\endpaper{\ifr@ffile
\immediate\write16{Cross References written on []\jobname.REF.}\fi\r@fend}

\catcode`@=12

\citeall\refto		

\input epsf
\parindent=20pt
\magnification 1200
\vsize=8.0in
\hsize=5in
\voffset=0.0in
\hoffset=0.0in

\baselineskip 12pt plus 0pt minus 0pt

\font\rmmthree=cmbx10 scaled 1500
\font\rmmtwo=cmbx10 scaled 1200
\font\rmmoneB=cmbx10 scaled 1100

\font\rmmoneI=cmti10 scaled 1100

\font\ninerm=cmr10 scaled 900

\font\eightit=cmti10 scaled 800
\font\eightrm=cmr10 scaled 800
\font\eightbf=cmbx10 scaled 800

\newcount\eqnumber

\def\chapbegin#1{\centerline{\rmmthree #1}\nobreak\bigskip
\eqnumber=1}

\def\sectbegin#1#2{\bigskip\bigskip\bigbreak\leftline{\rmmtwo #1 #2}\nobreak
    \medskip\nobreak \def\sectno{#1} \eqnumber=1 \mark{#1 \enskip#2}}

\def\nosectbegin#1{\bigskip\bigbreak\leftline{\rmmtwo #1}\nobreak\medskip}

\def\subbegin#1{\bigskip\medbreak\leftline{\bf #1}\nobreak
    \medskip\nobreak}

\def\subhead#1{\smallskip\goodbreak\noindent{\it #1}\nobreak\vskip 0pt\nobreak}


\def\new{{\sectno.\the\eqnumber}\global\advance\eqnumber by 1}
\def\delaynew{{\the\eqnumber}}
\def\nownew{\global\advance\eqnumber by 1}
\def\last{\advance\eqnumber by -1 {\the\eqnumber}
    \global\advance\eqnumber by 1}
\def\eqnam#1{
\xdef#1{\sectno .\the\eqnumber}}



\def\caption#1#2{
\baselineskip 10pt\noindent\narrower\rm\hbox{\eightbf
#1}:\quad\eightrm
#2 \smallskip}

\def\picture #1 by #2 (#3){
  \vbox to #2{
    \hrule width #1 height 0pt depth 0pt
    \vfill
    \special{picture #3} 
    }
  }

\def\scaledpicture #1 by #2 (#3 scaled #4){{
  \dimen0=#1 \dimen1=#2
  \divide\dimen0 by 1000 \multiply\dimen0 by #4
  \divide\dimen1 by 1000 \multiply\dimen1 by #4
  \picture \dimen0 by \dimen1 (#3 scaled #4)}
  }


\chapbegin{Quantitative String Evolution} \vskip 10pt
\centerline{\rmmoneB C.$\,$J.$\,$A.$\,$P. Martins\footnote{*}{\noindent \eightrm
Also at C. A. U. P., Rua do Campo Alegre 823, 4150 Porto,
Portugal} {\rmmoneI ~and~} E.$\,$P.$\,$S. Shellard} \vskip 8pt
\centerline{\rmmoneI Department of Applied Mathematics and Theoretical Physics} 

\centerline{\rmmoneI University of Cambridge}

\centerline{{\rmmoneI Silver Street, Cambridge 
~CB3 9EW, UK}\footnote{$^\dagger$}{\noindent \eightrm{\eightit Email}: cjapm10$\,$@$\,$damtp.cam.ac.uk and
epss$\,$@$\,$damtp.cam.ac.uk
\vskip0pt 
Paper submitted to {\eightit Physical Review D}.\smallskip}}
\medskip
\medskip

\centerline{\rmmoneB Abstract}
\medskip
\baselineskip 11pt plus 1pt minus 1pt
{\narrower{\ninerm 
\noindent An analytic model of long string network evolution, recently developed by the authors,
is presented in detail, and modified to describe string loop evolution. By treating the average
string velocity, as well as the characteristic lengthscale, as dynamical variables, one can
include the effects of frictional forces on the evolution of the network.
This generalized `one-scale' model provides a quantitative picture of the
complete evolution of a string network, including the prediction of previously
unknown transient scaling regimes and a detailed analysis of the evolution of
the loop population. The evolution of all cosmologically interesting string
networks is studied in detail, and possible consequences of our results are
discussed.
 
\smallskip}}  

\sectbegin{1~}{Introduction}
\medskip
\nobreak

\noindent Symmetry breaking phase transitions in the early universe inevitably
produce topological defects of one form or another.  Cosmic strings are of particular 
interest in this context, unlike some other defects, because the evolution of a string
network does not dramatically alter the standard cosmology.  In fact, superheavy 
strings associated with a grand unification phase transition provide a much-studied model
for the initial fluctuations for galaxy formation, also leaving imprints in the 
cosmic microwave radiation background.  But cosmological interest in strings is not restricted
to GUT scales, since strings could have formed at lower energies such as 
electroweak or Peccei-Quinn symmetry breaking with potentially important consequences, 
respectively, for baryogenesis or dark matter (axions).
Before studying the astrophysical consequences of strings, however, one must know how they are
formed and how they evolve. Due to their statistical nature, the best analytic approach consists
of doing `string thermodynamics', that is, describing the string network by a small number of
averaged quantities.

The serious analytic study of cosmological string networks began one decade ago with 
Kibble's `one-scale' model\refto{k85} (later modified by Bennett\refto{b1}). In this work it
was assumed that the evolution of the long-string network could be described using a
single lengthscale, which is usually called the `correlation length'. One then
supposes that a scaling solution exists at late times and ends up showing that
such a solution will in fact exist and be stable subject to conditions on the
loop production mechanisms. Note that in this model it is conceivable that a
string network could dominate the energy density of the universe\refto{kd}.

A step forward in the understanding of string network evolution was
provided by numerical simulations (see for example\refto{bb}). In
short, these confirmed the large-scale features of Kibble's model, namely
regarding the existence and stability of the scaling solution (at least in the
radiation era), but also showed that it neglects important physical processes
on small scales. In
particular, simulations revealed the existence of a significant amount of
small-scale structure on long strings, with loops being predominantly produced
at the smallest scales that can be sampled numerically.  This change in the
understanding of the mechanism of loop formation had of course important
consequences, notably in the cosmic string scenario for galaxy and large-scale
structure formation.

These findings triggered new efforts on the analytical side to try to account for small-scale
structure.  Notably, Austin, Copeland and Kibble developed a model\refto{ack} where the evolution of
the network is described by three different lengthscales, one of which aims to explicitly describe
the presence of small-scale structure. This model also includes a very simple treatment of the
effects of gravitational radiation. Apart from confirming the predictions of the one-scale
model for the large-scale properties of the network, the main result of this model is the
suggestion that the effects of gravitational back-reaction are needed if this `third
lengthscale' is to reach scaling. A less attractive aspect of this model is that it has to resort
to an unappealingly large number of unspecified parameters. There have also been studies of the
evolution of the linear kink density in what is effectively a `one-scale model context' which 
anticipated these results by
Allen \& Caldwell\refto{aca} and later by Austin\refto{aus}.

However, it is usually said that it is very difficult to build a house if one starts with the
roof. It is therefore restrictive to try to build models whose only aim is to describe an
eventual linear scaling regime. In particular, there is a fundamental ingredient in the
evolution of a cosmic string network that has been neglected until very recently, namely
frictional forces due to particle--string scattering, which are important for some time after
the string-forming phase transition. It should be kept in mind that the period immediately
after string formation is by no means irrelevant, eg for baryogenesis mechanisms involving
cosmic strings. Furthermore, for electroweak strings the friction-dominated epoch only ends in
the matter era, lasting almost through to the present day.

A model of string network evolution including the effects of frictional forces has been 
recently proposed by the authors\refto{ms}. Although it should be seen as the basis for further work, the model is already predictive
enough to be testable in both numerical and laboratory experiments, if not cosmologically.

The model is a simple
generalization of the `one-scale' model in which the average string rms velocity becomes a
dynamical variable. At present, the model does not include small-scale structure (although
there are significant hints on how to do it) or other potentially important effects such as loop
reconnections onto the long string network. Nevertheless, it will be shown that this simple
model provides the first, fully quantitative description of the complete evolution of a string
network in the early universe (see sections 4-5).  In particular, the
existence of two different transient scaling regimes in the epoch of
friction-dominated dynamics 
are established (one of which was previously suggested by Kibble). Also, it is apparent in 
this
model that cosmic strings will never dominate the energy density of the universe (for reasons
other than the statistical physics arguments of Albrecht \& Turok\refto{albrecht}).

Cosmic string loops decay fairly quickly after their formation. For this reason, their
contribution to the seeding of gravitational instabilities or cosmic microwave background
anisotropies is  thought to be subdominant relative to long strings. This fact probably
explains why the evolution of the loop distribution has been comparatively neglected in the
literature. This gap can be filled by the model to be described here. With simple
modifications---the more important of them being the use of the physical loop size $\ell$
rather then the correlation length $L$---this model can also be used to study
the evolution of the loop distribution. In particular, it will be shown that,
depending on the parameters characterizing loop production and lifetimes, there
is more energy density in loops than in long strings.

Strings (and topological defects in general), of course, are not exclusive to the
early universe. They exist (and have been seen) in a wide variety of condensed
matter contexts, including metal crystallization\refto{metc}, liquid
crystals\refto{lctw,cdty}, superfluid helium-3\refto{msgv} and
helium-4\refto{zurek}, and superconductivity\refto{abrk}. Our generalized
`one-scale' model can also be used to describe vortex-string evolution in
condensed matter contexts (with advantages over previously used approaches).
In particular, some well-known results can be readily reproduced, and new
quantitative predictions regarding loop production can be made. These issues
are discussed in a companion paper\refto{ms2}.

The structure of this paper is as follows. In the next section, after a short
review of string dynamics, the evolution equations for the `characteristic
lengthscale' and the average velocity of the long string network and each
individual loop are derived and justified. The cases of strings arising from
the breaking of gauge and global symmetries are both considered. The validity
of these `averaged' evolution equations is then tested against simple loop solutions in section
3. Section 4 contains a discussion on the importance of the friction force in the early universe,
together with the analysis of the different scaling laws in the model for both
long strings and loops. There are also some preliminary comparisons with
numerical simulations. We then proceed to a detailed and individual
analysis of the four physically relevant cases: gauge electroweak and GUT, and global
axionic and GUT strings are the subject of section 5. Finally, section 6
contains conclusions.

Throughout this paper we will use fundamental units in which
$\hbar=c=k_B=1$.

\sectbegin{2}{A generalized `one-scale' model}
\bigskip
\nobreak

\subbegin{A. String dynamics with friction}
\bigskip

\noindent A string sweeps out a  two-dimensional surface (the worldsheet) which can be described
by spacetime coordinates
$x^{\mu}$ and worldsheet coordinates $\sigma^a$, $x^{\mu}=x^{\mu}(\sigma^a)$ ; the line
element is then
\eqnam{\strlel}
$$
ds^2=g_{\mu\nu} \, x^{\mu}_{,a} \, x^{\nu}_{,b} \, d\sigma^a \,
d\sigma^b = \gamma_{ab} \, d\sigma^a \, d\sigma^b \, , \eqno (\new)
$$
where $g_{\mu\nu}$ and $\gamma_{ab}$ are respec\-tively the four-dimen\-sional spacetime and
two-dimen\-sional string worldsheet metrics. For the case of a gauge (global) string, one can then
derive the Nambu (Kalb-Ramond) action from the abelian-Higgs (Goldstone) model
on the assumption that the scale of perturbations along the string is much
larger than its width $\delta$. (In the global case, one also makes use of the
equivalence between a real massless scalar field and a two-index antisymmetric
tensor field.) One finds
\eqnam{\actions}
$$
S=\cases{\mu\int\sqrt{-\gamma} d\sigma^2,&Gauge \cr 
\mu_o\int\sqrt{-\gamma} d\sigma^2 + {1\over 6}\int\sqrt{-g}H^2d^4x + 2\pi\eta\int
B_{\mu\nu}d\sigma^{\mu\nu},&Global \cr} \, , \eqno (\new)
$$
where $B_{\mu\mu}$ is the antisymmetric tensor field, $H_{\mu\nu\lambda}$ is its field strength
and $d\sigma^{\mu\nu}$ is the worldsheet area element. Hence the Nambu action is proportional
to the area swept out by the string. By varying this action one obtains the following equations
of motion
\eqnam{\nofri}
$$
x^{\nu}{}_{,a}{}^{;a}+\Gamma^{\nu}_{\tau\lambda} \, \gamma^{ab} \,
x^{\tau}{}_{,a} \, x^{\lambda}_{,b}=\cases{0,&Gauge \cr 
{2\pi\eta\over\mu_o} H^{\nu}_{\tau\lambda} \, \epsilon^{ab} \,
x^{\tau}{}_{,a} \, x^{\lambda}_{,b} ,&Global \cr} \, . \eqno (\new)
$$
It should be noted that in the global case $\mu_o$ is the `bare' (unrenormalized) energy per
unit length. However, it can be shown that if one distinguishes between the external and
self-field contributions to
${\bf H}$ and sets ${\bf H}_{self}=0$ the above equations still hold with $\mu_o$ replaced by the
renormalized energy per unit length, denoted by $\mu$\refto{dsmag}.

Still, a crucial ingredient for string evolution is missing. Since strings move through
a background radiation fluid, their motion is  retarded by particle scattering. Vilenkin
has shown\refto{v1} that this effect can be described by a frictional force per unit length that
can be written
\eqnam{\ff}
$$
{\bf F}_{\rm f}=-{\mu\over \ell_{\rm f}}{{\bf v}\over \sqrt{1-v^2}} \, ,\eqno (\new)
$$
where ${\bf v}$ is the string velocity and $\ell_{\rm f}$ will be called the `friction
lengthscale'; its explicit value depends on the type of symmetry involved. For a gauge
string, the main contribution comes from Aharonov-Bohm scattering\refto{rsa}, while in the global
case it comes from Everett scattering\refto{epa}. Then we respectively have
\eqnam{\frilen}
$$
\ell_{\rm f}=\cases{{\mu\over\beta T^3},&Gauge \cr {\mu\over\beta T^3} \ln^2(T\delta),&Global \cr}
\, ,\eqno (\new)
$$
where $T$ is the background temperature and $\beta$ is
a numerical factor related to the number of particle species interacting with the string (strictly
speaking, its value is slightly different in the two cases, but a common symbol will be used
for simplicity). Specifically in the gauge case we have
\eqnam{\expb}
$$
\beta={2\zeta (3)\over \pi^2}\sum_a b_a sin^2(\pi\nu_a) \, ,\eqno (\new)
$$
where this sum is taken over effectively massless degrees of freedom, $\nu_a$ is the phase change
experienced by a particle transported around the string and $b_a$ is $1$ for bosons and $3/4$ for
fermions. Hence Aharonov-Bohm scattering will occur for particles with non-integer $\nu's$;
see the paper by Alford \& Wilczek in\refto{rsa} for an example of a model with such values. It
should also be noted that the Everett scattering formula is only valid when the particle wavelength
is much larger than the string thickness $\delta$.

It is then straightforward to show that the frictional force per unit length (\ff) can be
included in the equations of motion (\nofri) by adding the term
\eqnam{\newt}
$$
\left(U^{\nu}-x^{\nu}_{,a}x^{\sigma,a}U_{\sigma}\right){1\over\ell_{\rm f}} \, , \eqno (\new)
$$
($U^{\nu}$ being the four-velocity of the background fluid) on its right-hand side.

Now consider string motion in an FRW universe with the line element, 
\eqnam{\lel} 
$$
ds^2=a^2(\tau) \, \left(d\tau^2-{\bf dx}^2\right) \, ; \eqno (\new)
$$ 
then $U^{\nu}=\left(a^{-1},{\bf 0}\right)$ and choosing the gauge conditions $\sigma=\tau$ (ie,
identifying conformal and worldsheet times) and $\dot{\bf x}\cdot{\bf x}'=0$ (ie, imposing that the
string velocity be orthogonal to the string direction) the string equations of
motion with the frictional force (\ff) in the background (\lel)  can then be expressed
as\refto{v1,tb}
\eqnam{\spc}
$$
{\ddot{\bf x}}+\left(2{\dot a\over a}+{a\over \ell_{\rm f}}\right)\left(1-{\dot{\bf
x}}^2\right)\dot{\bf x} ={1\over \epsilon}\left({{\bf x}'\over \epsilon}\right)'
\, , \eqno (\new)
$$
\eqnam{\tc}
$$
{\dot\epsilon}+\left(2{\dot a\over a}+{a\over\ \ell_{\rm f}}\right){\dot{\bf x}}^2\epsilon=0
\, , \eqno (\new)
$$
where the `coordinate energy per unit length' $\epsilon$ is defined by
\eqnam{\epsil}
$$
{\epsilon^2=}{{{\bf x}'}^2\over {1-{\dot{\bf x}}^2}} \, , \eqno (\new)
$$
and dots and primes respectively denote derivatives with respect to $\tau$ and
$\sigma$.
\footnote{${}^1$}{\noindent \eightrm Note that reparametrizations of $\sigma$
can be absorbed into changes of
$\epsilon$.} This form of the evolution equations  proves to be particularly useful because
dissipation is naturally incorporated in the decay of the coordinate energy density $\epsilon$,
while preserving the gauge conditions.  

Incidentally, it has been shown\refto{dsmag} that a global string will behave as a superfluid
vortex if it is introduced in a homogeneous background of the form
\eqnam{\hext}
$$
H_{ext}^{ijk}=\sqrt{\rho_h}\epsilon^{ijk} \,  \eqno (\new)
$$
(physically, this corresponds to giving it angular momentum). The interaction between this
background and the string gives rise to an additional force, known as the (relativistic) Magnus
force, and (\spc) becomes
\eqnam{\magnus}
$$
{\ddot{\bf x}}+\left(2{\dot a\over a}+{a\over \ell_{\rm f}}\right)\left(1-{\dot{\bf
x}}^2\right)\dot{\bf x} ={1\over \epsilon}\left({{\bf x}'\over \epsilon}\right)' + {1\over\epsilon}
{\rho_h\over\mu}\dot{\bf x}\wedge{\bf m} \, , \eqno (\new)
$$
where
\eqnam{\circ}
$$
{\bf m}={4\pi\eta\over\sqrt{\rho_h}}{\bf x}' \,  \eqno (\new)
$$
is the circulation vector; the energy equation (\tc) remains unchanged.

\subbegin{B. Lengthscale evolution}
\bigskip

\noindent We can now proceed to average the string equations of motion to  describe the
large-scale evolution of the string network. We therefore define the total string energy and the
average rms string velocity to be
\eqnam{\et}
$$
E=\mu a(\tau)\int\epsilon d\sigma \,, \eqno (\new)
$$
\eqnam{\vv}
$$
v^2\equiv \langle{\dot{\bf x}}^2\rangle={{\int{\dot{\bf x}}^2\epsilon d\sigma}\over{\int\epsilon
d\sigma}} \, .\eqno (\new)
$$
Differentiating (\et) and using (\tc) and (\vv), we see that 
the total string energy density $\rho \propto E/a^3$ will obey the following evolution equation
(in terms of physical time $t$):
\eqnam{\dens} 
$$
{d\rho\over dt}+\left(2H\left(1+v^2\right)+{v^2\over \ell_{\rm f}}\right)\rho=0 \, . \eqno (\new)
$$

Equation (\dens) incorporates both long strings and small,
short-lived loops which have (in general) a low probability of interacting with other strings before
their demise. We shall study the evolution of the long-string network on the assumption that
it can be characterized by a single lengthscale $L$; this can be interpreted as the inter-string
distance or the `correlation length'. Strings larger than $L$ will be called long or `infinite';
otherwise they will be called loops. For Brownian long strings, we can define the `correlation
length' $L$ in terms of the network density\footnote{${}^2$}{\noindent \eightrm
Throughout this paper the subscript `$\infty$' refers to properties of the long (`infinite')
string network.} $\rho_\infty$ as
\eqnam{\correlation} 
$$
\rho_{\infty}\equiv{\mu\over L^2}\,.\eqno(\new)
$$

Following Kibble\refto{k85}, the rate of loop production from long-string collisions can be
estimated as follows. Conceptually, we divide the network into a collection of segments of length
$L$, each in a volume $L^3$.  Consider another segment of length $l$ moving with a velocity
$v_\infty$; the probability of it encountering one of the other segments within a time $\delta
t$ is approximately
$lv_\infty\delta t/ L^2$. Consistently with our `one-scale' assumption, we then assume that the
probability of such an intersection creating a loop of length in the range $l$ to $l+dl$ will
be given by a scale-invariant function $w\left({l/L}\right)$. The rate of energy loss into
loops is then given by
\eqnam{\rtl}
$$
\left({d\rho_{\infty}\over dt}\right)_{\rm to\ loops}=\rho_{\infty}{v_\infty\over
L}\int w\left({\ell\over L}\right){\ell\over L}{d\ell\over L}\equiv {\tilde
c}v_\infty{\rho_{\infty}\over L} \, ,
\eqno (\new)
$$
where  the loop `chopping' efficiency ${\tilde c}$ is assumed to be constant.
Note that in previous
analyses without friction $v_\infty$ was assumed to be constant and absorbed into the
definition of ${\tilde c}$.

Finally, by subtracting the loop energy losses (\rtl) from (\dens) and then using
(\correlation),  we obtain the overall evolution equation for the characteristic lengthscale
$L$, 
\eqnam{\evl} 
$$
2 {dL\over dt}=2HL(1+v_\infty^2)+{Lv_\infty^2\over \ell_{\rm f}}+{\tilde c}v_\infty \, . \eqno
(\new)
$$
Note that with the exception of the expansion term, all terms on the right-hand side are
velocity-dependent; this will have important consequences (see below).

\subbegin{C. Loop evolution}
\bigskip

\noindent On the other hand, we can also study the evolution of the loop density and
distribution. The traditional approach is to define
$n_\ell(\ell,t)d\ell$ to be the number density of loops with length in the range $(\ell,\ell
+d\ell)$ at time $t$; the corresponding loop energy density distribution is
\eqnam{\ledd} 
$$
\rho_{\ell}(\ell,t)d\ell = \mu\ell n_\ell(\ell,t)d\ell \, . \eqno (\new)
$$
Note that the total loop energy density is
\eqnam{\ledd} 
$$
\rho_{o}\equiv\int\rho_{\ell}(\ell,t)d\ell \, , \eqno (\new)
$$
and $\rho=\rho_{\infty}+\rho_{o}$\footnote{${}^3$}{\noindent \eightrm
Throughout this paper the subscript `$o$' refers to properties of the entire loop
population; `$\ell$' refers to the loops with length in the range $(\ell,\ell +d\ell)$.} From
our assumptions on the loop production rate (\rtl) it is then easy to see that
\eqnam{\evledd} 
$$
{d\rho_{\ell}\over dt}+\left(2H\left(1+v^2_{\ell}\right)+{v_{\ell}^2\over \ell_{\rm
f}}\right)\rho_{\ell}=g\mu{v_\infty\ell\over L^5}w\left({\ell\over L}\right) \, , \eqno (\new)
$$
where $g$ is a Lorentz factor accounting for the fact that loops are created with non-zero
center-of-mass kinetic energy (lost through velocity redshift).  However, note that this
equation is `static', in the sense that it does not include loop decay mechanisms (eg, via the
emission of gravitational, Goldstone boson or electromagnetic radiation, as the case may be).

Instead, we start by using our analytic model to describe the evolution of each individual loop.
Knowing the energy density transferred from long strings into loops  and estimating their sizes
at formation (see below), one can numerically determine the energy density in loops and other
relevant quantities at all times. This formalism is that we does not allow for
loop reconnections, which are unimportant for GUT strings (but can be relevant for high-density
electroweak or axionic string networks---see section 5); furthermore
self-intersections could be included by carefully defining an `effective' loop production size.

The physical size of a loop is simply given by
\eqnam{\physl}
$$
\ell=a(\tau)\int_{loop}\epsilon d\sigma \, ; \eqno (\new)
$$
its time derivative can be easily calculated using (\tc). However one must still subtract
energy (hence length) losses due to radiative processes. For the case of a gauge string, this
can be roughly estimated from the quadrupole formula
\eqnam{\qdp}
$$
\left({dE\over dt}\right)_{rad}\sim G\left({d^3D\over dt^3}\right)^2\sim G\mu^2v^6 \, ,
\eqno (\new)
$$
($D\sim\mu \ell^3$ being the loop's quadrupole moment). Again, note that loop velocity is
usually assumed constant ($v_o^2=1/2$) and not included in (\qdp). (This is obviously correct
in the `free' regime, but it is not a good assumption in the friction-dominated regime.)
Then we define
\eqnam{\lrad}
$$
\left({d\ell\over dt}\right)_{rad}\equiv - \Gamma'G\mu v^6 \, , \eqno (\new)
$$
where according to numerical estimates $\Gamma'\sim8\times65$ (note that the original
parameter $\Gamma$ was calculated in flat space, where $v_o^2=1/2$). Then the evolution
equation for the physical loop size has the form
\eqnam{\evlps}
$$
{d\ell\over dt}=(1-2v^2_\ell)H\ell -{\ell v^2_\ell\over\ell_f} - \Gamma'G\mu v^6_\ell \, .
\eqno (\new)
$$
Again, all but the expansion term are velocity-dependent.

For the case of axionic strings, however the emission of gravitational is
subdominant with respect to the emission of axions. The above expressions will
still be valid with the replacement
\eqnam{\axcase}
$$
\Gamma'G\mu\longrightarrow
q\equiv{\Gamma'\over2\pi}{1\over\ln\left({L\over\delta}\right)}
\, ,
\eqno (\new)
$$
where $\delta$ is the string thickness.

Now, we will assume that loop production is `monochromatic', ie that loops formed at a time
$t_p$ have an initial length
\eqnam{\inls}
$$
\ell(t_p)=\alpha(t_p)\ L(t_p) \, .
\eqno (\new)
$$
Notice that we are implicitly saying that the loop size at formation depends both on the
large-scale properties of the network (through the correlation length) and on the small-scale
structure it contains (through the parameter $\alpha$). At this stage, since the model does
not include small-scale structure, we shall resort to physical arguments about radiative 
backreaction etc.\ to
obtain an ansatz for $\alpha$ when necessary.

With this ansatz the scale-invariant loop production function $w$ (see (\rtl)) becomes
\eqnam{\wansatz}
$$
w\left({\ell\over L}\right)={{\tilde c}\over\alpha}\delta\left({\ell\over L}-\alpha\right) \, ,
\eqno (\new)$$
and the rate of energy loss into loops becomes
\eqnam{\rtlalpha}
$$
\left({d\rho_{\infty}\over dt}\right)_{\rm to\ loops}=g\mu{\tilde
c}{v_\infty\over L^3} \, ,
\eqno (\new)
$$
with $g$ being the Lorentz factor as above.

Hence the energy density converted into loops from time $t$ to $t+dt$ is
\eqnam{\rozdt}
$$
d\rho_o(t)=g\mu{\tilde
c}{v_\infty\over L^3}dt \, ;
\eqno (\new)
$$
this corresponds to a fraction
\eqnam{\rozfrdt}
$$
{d\rho_o(t)\over\rho_{\infty}(t)}=g{\tilde c}{v_\infty\over L}dt \, 
\eqno (\new)
$$
of the energy density in the form of long strings at time $t$. Then using our ansatz
(\wansatz), the corresponding number of loops produced in a volume $V$ is
\eqnam{\numlo}
$$
dN(t)=g{{\tilde c}\over\alpha}{v_\infty\over L^4}Vdt \, ; \eqno (\new)
$$
hence the ratio of the energy densities in loops and long strings at time $t$ is 
\eqnam{\ratio}
$$
\varrho(t)\equiv{\rho_o(t)\over\rho_{\infty}(t)}=gL^2(t)\int_{t_c}^{t}{dN(t')\ell(t,t')\over
V}=g{\tilde c}L^2(t)\int_{t_c}^{t}{v_\infty (t')\over L^4(t')}{\ell(t,t')\over\alpha (t')}dt'
\, ,
\eqno (\new)
$$
where $t_c$ is the moment of the network formation and $\ell(t,t')$ is the length at time $t$
of loops produced at time $t'$.  Here, we neglect
the initial loop population at $t=t_c$; this probably has a scale-invariant form (see the 
discussion in ref.\refto{VV}.

We can therefore numerically (and, in some simple limit
cases, analytically) determine the loop density at all times.  This generalized `one-scale'
model can therefore provide a complete description of a string network.

\subbegin{D. Velocity evolution}
\bigskip

\noindent We now consider the evolution of the average string velocity $v$. A
non-relativ\-istic equation can be easily obtained: it is just Newton's law,
\eqnam{\nonrelvel}
$$
\mu{dv\over dt}={\mu\over R}-\mu v\left(2H+{1\over \ell_{\rm f}}\right) \, .\eqno(\new)
$$
This merely states that curvature accelerates the strings while damping (both from friction and
expansion) slows them down. On dimensional grounds, the force per unit length due to curvature should
be $\mu$ over the curvature radius $R$. The form of the damping force can be found
similarly. 

A relativistic
generalization of the velocity evolution equation (\nonrelvel) can be obtained more rigorously
by differentiating (\vv):
\eqnam{\evv}
$$
{dv\over dt}=\left(1-v^2\right)\left[{k\over R}-v\left(2H+{1\over \ell_{\rm f}}\right)\right]
\, . \eqno (\new)
$$
This is exact up to second-order terms. To obtain the damping term we have taken
$\langle{\dot{\bf x}}^4\rangle=\langle{\dot{\bf x}}^2\rangle^2$. Writing
${\dot{\bf x}}^2=(1+{\bf p}\cdot {\bf q})/2$ (${\bf p}$ and ${\bf p}$ being unit left- and
right-movers along the string) and defining $\varsigma\equiv -\langle{\bf p}\cdot {\bf q}\rangle$
the difference between the two is
\eqnam{\difxdots}
$$
\langle{\dot{\bf x}}^4\rangle - \langle{\dot{\bf x}}^2\rangle^2 ={1\over4}
\left[\langle({\bf p}\cdot {\bf q})^2\rangle-\varsigma^2\right] \, . \eqno (\new)
$$
Note that numerical simulations of string evolution indicate that $\varsigma_{rad}\sim 0.14$ and
$\varsigma_{mat}\sim 0.26$, so this difference should also be small.
As for the curvature term, we have
introduced $R$ via the definition of the curvature radius vector,
\eqnam{\crv}
$$
{a(\tau)\over R}\hat{\bf u}={d^2{\bf x}\over ds^2} \, , \eqno (\new)
$$
where $\hat{\bf u}$ is a unit vector and $s$ is the physical length along the string (related to
the coordinate length $\sigma$ by $ds=|{\bf x}'|d\sigma = \left(1-{\dot{\bf x}}^2\right)^{1/2}
\epsilon d\sigma$).  The dimensionless parameter $k$ is defined by
\eqnam{\dfk}
$$
\langle(1-{\dot{\bf x}}^2)({\dot{\bf x}}\cdot\hat{\bf u})\rangle\equiv kv(1-v^2) \, . \eqno (\new)
$$
The parameter $k$ is related to the presence of small-scale structure on strings: on a
perfectly smooth string, $\hat{\bf u}$ and ${\dot{\bf x}}$ will be parallel so $k=1$ (up to a
second-order term as above), but this need not be so for a wiggly string. On the other hand,
$k=0$ for a string loop in flat spacetime---in particular, it is easy (although rather
tedious) to show using its definition (\dfk) that this is indeed the case for all Kibble-Turok
and all Burden loops. In that sense, flat spacetime is the case of maximal small-scale
structure; this is not surprising, since there is no mechanism available for kink
decay. In a model including small-scale structure, $k$ would probably be a further dynamical
variable. For this model, however, we shall use an ansatz interpolating between these two
extreme regimes motivated by comparisons with microscopic evolution.

Consider a particular string loop.
On large enough scales, they are `frozen' in the background, being conformally
stretched by expansion (note that very large loops should only form in friction-dominated
regimes). Then they should have relatively little small-scale structure, and taking
$k\approx1$ should be a good assumption. On the other hand, on small enough scales strings
behave as if they were in flat spacetime, so one requires that $k\rightarrow0$ as
$R\rightarrow0$. In particular, in the case of loops, demanding that their limiting velocity be
$v^2_\ell=1/2$ leads to the requirement that $k\propto R$ as $R\rightarrow0$. The remaining
point consists in noting that what is dynamically meant by `large' and `small' scales depends
not on the relative size of
$R$ and the horizon but on the relative size of $R$ and the `damping length' defined as
\eqnam{\daml}
$$
{1\over\ell_d}\equiv2H +{1\over\ell_f} \, . \eqno (\new)
$$
With these requirements in mind, and after comparing with the `microscopic' (ie, unaveraged)
evolution of some simple solutions (to be described in the next section) one arrives at the
following ansatz:
\eqnam{\kans}
$$
k=\cases{1,&${R\over\ell_d}>\chi$ \cr {1\over\sqrt2}{R\over\ell_d},
&${R\over\ell_d}<\chi$ \cr} \, , \eqno (\new)
$$
where $\chi$ is a numerical coefficient\refto{gs} of order unity.
Recall that the physical loop length is approximately $\ell=2\pi R$; since we will be
considering each individual loop, this ansatz applies immediately in that case.

For the case of the long-string network (for which $L=R$), the reasoning is
roughly the same in the regime where $L\gg\ell_d$. However, the opposite regime
never arises (at least, in the early universe), so there is no simple way of
inferring the
$k$ behaviour. The closest a network gets is in the linear scaling regime in the early
universe, where the ratio is less than, but still of order unity. Of course there is also the
problem that $k$ has a slightly different definition---namely, it is an average over the whole
network, on a scale of the `correlation length' $L$. This makes its physical interpretation
slightly less clear. For these reasons, when discussing this regime we will start by assuming
that
$k$ is a constant of order unity. Further justification for this assumption, and a discussion
of the possible use of (\kans) or other alternatives can be found in section
4.

Equations (\evl), (\evlps) and (\evv) form the basis of our generalized `one-scale' model, which
we will now proceed to apply in several different contexts.  We note that the
velocity-independent `one-scale' model (\evl) has proved to be successful in describing the
large-scale properties of cosmic string networks in numerical simulations. Any deficiencies
seem to be associated with the emergence of significant small-scale structure, that is,
propagating kinks and wiggles on scales well below $L$.  In friction-dominated regimes,
therefore, we should anticipate improved quantitative agreement because of the suppression of
this substructure. 

\sectbegin{3}{`Averaged' versus `microscopic' evolution}
\bigskip
\nobreak

\noindent In order to check the validity of our `averaged' evolution model, and in particular
our ansatz for {\it k}, we will test it against simple loop solutions.

Firstly, consider a circular loop in flat spacetime but with a constant (non-infinite)
friction length---ie, a condensed-matter-like situation. We can describe the loop trajectory
simply by
\eqnam{\ccl}
$$
{\bf x}=r(\tau) (\sin\theta,\cos\theta,0) \,\qquad \theta\in[0,2\pi] \, . \eqno (\new)
$$
Then equations (\spc,\tc) reduce to\footnote{${}^4$}{\noindent \eightrm
Overdots denote differentiation with respect to conformal time
$\tau$, rather than physical time $t$.  In flat space, 
$\tau=t$.}
\eqnam{\smallr}
$$
{\ddot r}+(1-{\dot r}^2)\left({{\dot r}\over\ell_f}+{1\over r}\right)=0 \, . \eqno (\new)
$$

\midinsert
\vbox{\centerline{
\epsfxsize=.7\hsize\epsfbox{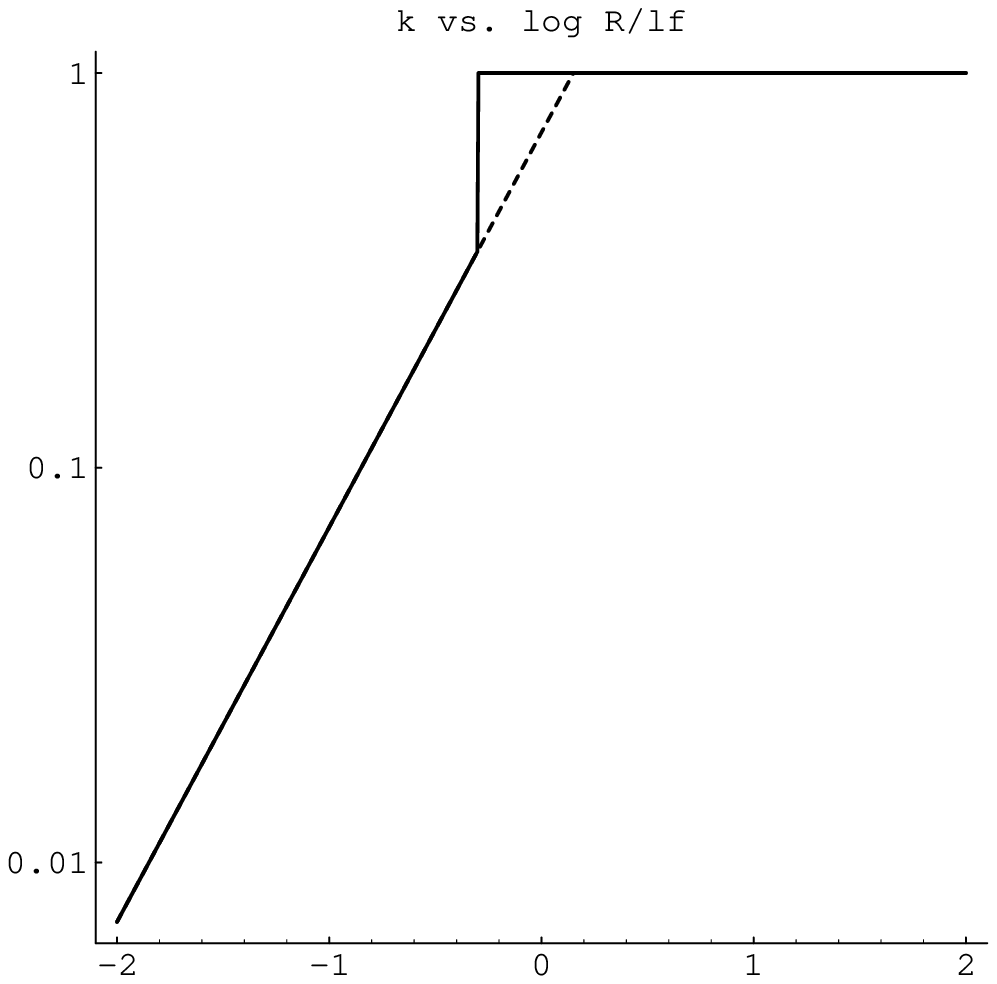}}
\vskip.5in
{\baselineskip 10pt\noindent\narrower\rm\hbox{\eightbf
Figure 3.1}:\quad\eightrm
{Two possible ansatzes for the parameter $k$
defined in (\dfk), specifically for strings
in a condensed-matter context as a function of the rescaled loop radius $R/\ell_f$.
For the case of string loops, the behaviour at large and small scales has a physical
justification (see text); numerical matching was improved for circular loops by
altering the transition point from 
$\chi=\sqrt{2}$ (solid curve) to $\chi\sim0.57$ (dashed curve). However, the ansatz
corresponding to the dashed curve might be relevant for long strings (see
section 4).}\smallskip}}
\endinsert

\noindent Note that the physical (`invariant') loop radius is
$R=r/\sqrt{1-{\dot r}^2}$, obeying
\eqnam{\bigr}
$$
{\dot R}=-{\dot r}^2{R\over\ell_f} \, ; \eqno (\new)
$$
also the `microscopic' velocity is $v=-{\dot r}$ and obeys
\eqnam{\unavv}
$$
{\dot v}=(1-v^2)\left({1\over r}-{v\over\ell_f}\right) \, . \eqno (\new)
$$

On the other hand, our averaged evolution equations (\evlps,\evv) take
the form\footnote{${}^5$}{\noindent \eightrm In this section, averaged
quantities will be denoted by overbars.}
\eqnam{\avev}
$$
{d{\bar R}\over dt}=-{\bar v}^2{{\bar R}\over\ell_f} \, , \qquad 
{d{\bar v}\over dt}=(1-{\bar v}^2)\left({k({\bar R})\over {\bar R}}-{{\bar v}\over\ell_f}\right)
\, . \eqno (\new)
$$
Notice the similarity between the two approaches. Loops with size much larger than the friction
length $\ell_f$ will be overdamped, with the velocity being approximately given by
\eqnam{\fovd}
$$
v\sim{\ell_f\over r} \, . \eqno (\new)
$$

\midinsert
\vbox{\centerline{%
\hskip4em\epsfxsize=.6\hsize\epsfbox{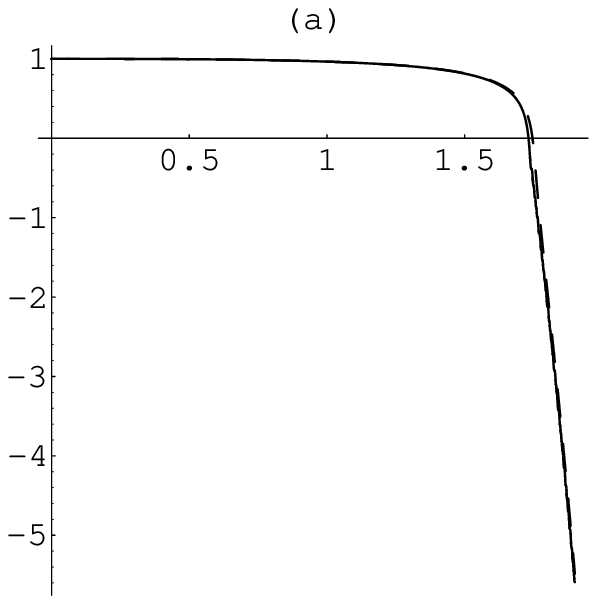}\hskip-10em\epsfxsize=.6\hsize\epsfbox{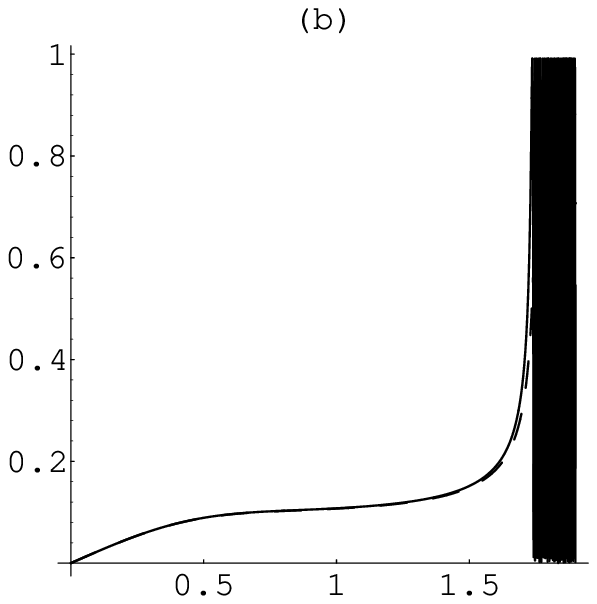}}
\vskip-.75in}
\endinsert
\midinsert
\vbox{\centerline{%
\hskip4em\epsfxsize=.6\hsize\epsfbox{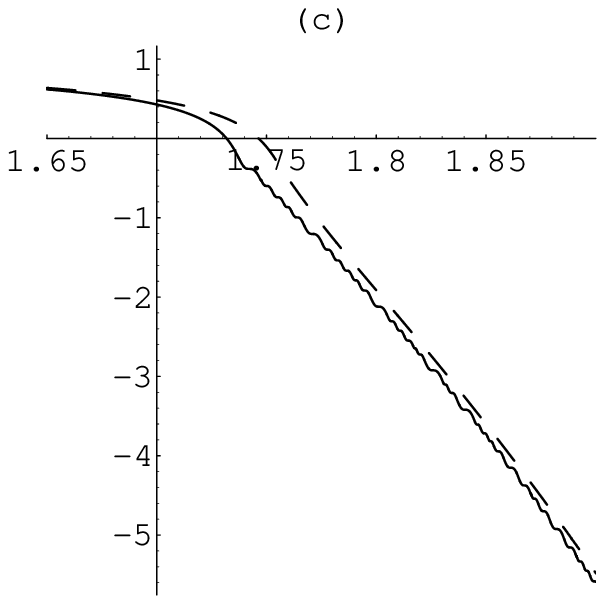}\hskip-10em\epsfxsize=.6\hsize\epsfbox{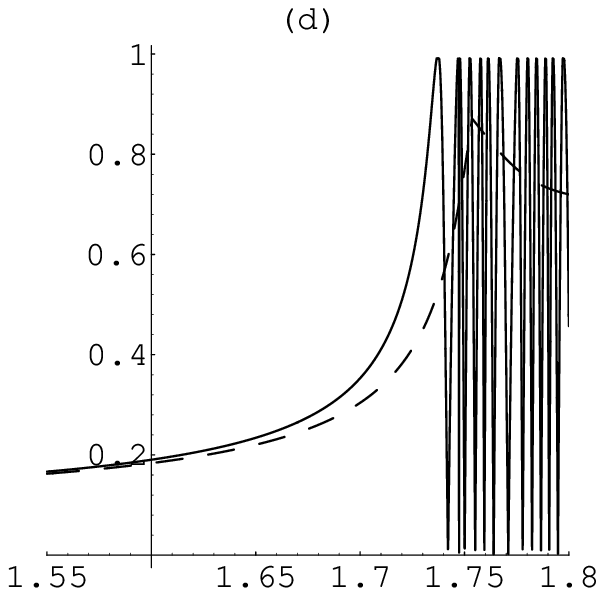}}
\vskip-.75in
{\baselineskip 10pt\noindent\narrower\rm\hbox{\eightbf
Figure 3.2}:\quad\eightrm
{Comparing the `microscopic'
(solid lines) and `averaged' (dashed) evolution equations for a circular loop in a
condensed-matter context. Length and time are in units of $\ell_f$, and the time axis is with a
logarithmic scale. Plot (a) depicts the log of the (rescaled) radius, while (b) depicts the
loop velocity.  The lower graphs are close-ups of the upper ones.}\smallskip}}
\endinsert

In this case the two sets of evolution equations actually coincide---hence justifying our
$k=1$ ansatz for large R. As the loop gains velocity $r$ and $R$ become significantly
different and this equivalence ceases to be valid. When $R$ becomes much smaller than
$\ell_f$, the loop still looses energy due to friction, but this is no longer effective in
damping its motion---the loop now begins to oscillate relativistically. In particular, over
one `period' $v$ oscillates between $0$ and
$1$ (ignoring nonlinear effects near $R=0$ due to the finite string width). But we know that the
averaged velocity should be
${\bar v}^2=1/2$ (in the small-scale limit); this is the physical reason why we need  $k$ to be a
`phenomenological' variable. As we mentioned previously, this requirement
fixes the behaviour of $k$ on small scales to be as shown in (\kans). The remaining question is
then how to match the two regimes.

\midinsert
\vbox{\centerline{%
\hskip4em\epsfxsize=.6\hsize\epsfbox{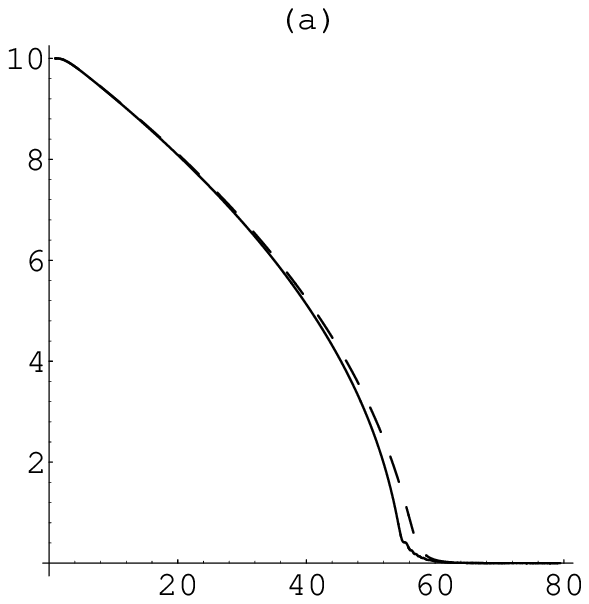}\hskip-10em\epsfxsize=.6\hsize\epsfbox{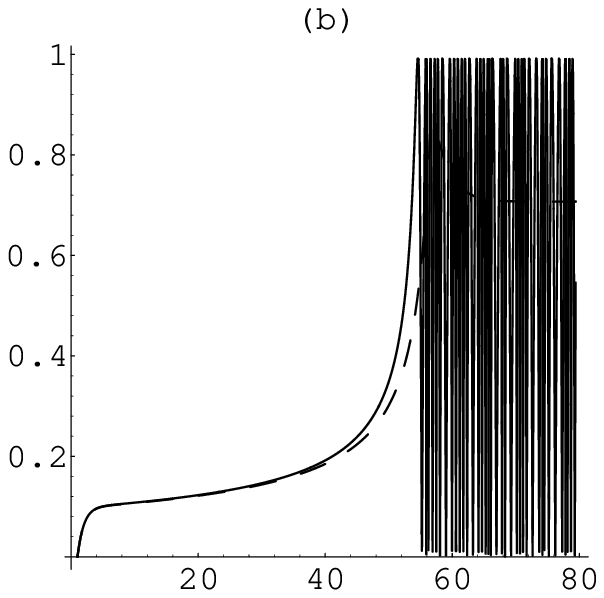}}
\vskip-.75in}
\endinsert
\midinsert
\vbox{\centerline{%
\hskip4em\epsfxsize=.6\hsize\epsfbox{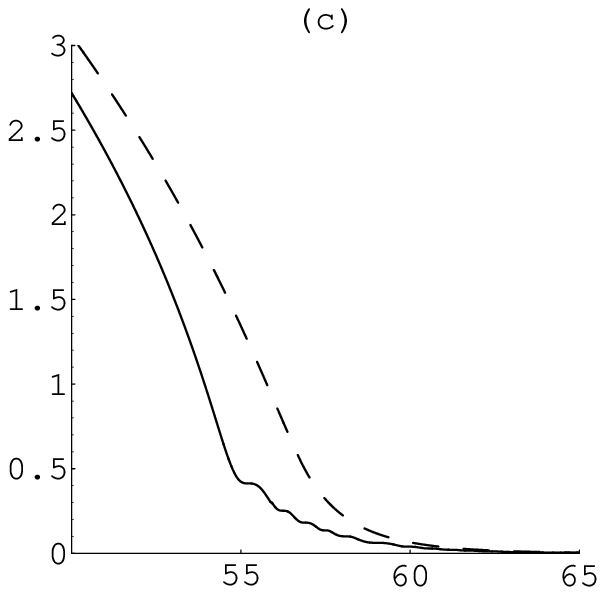}\hskip-10em\epsfxsize=.6\hsize\epsfbox{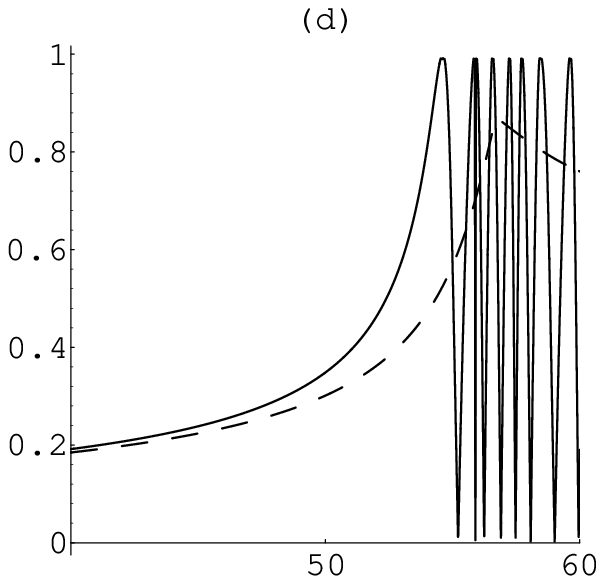}}
\vskip-.75in
{\baselineskip 10pt\noindent\narrower\rm\hbox{\eightbf
Figure 3.3}:\quad\eightrm
{Comparing the `microscopic'
(solid lines) and `averaged' (dashed) evolution equations for a circular loop in a
condensed-matter context. Length and time are in units of $\ell_f$, and the time axis is with a
linear scale. Plot (a) depicts the log of the (rescaled) radius, while (b) depicts the
loop velocity.  The lower graphs are close-ups of the upper ones.}\smallskip}}
\endinsert

First of all, we need a clear idea of when (and where) the transition occurs. A good guess
would be the moment of the `first collapse', ie, the moment when we first have $v=1$. In fact,
this turns out to be a well-defined event. As was first pointed out by Garriga and
Sakellariadou\refto{gs} (and can be easily seen by analytical or numerical study of the
equation of motion (\smallr)), circular loops with initial radius much larger than the friction
length always reach $v=1$ for the first time when
\eqnam{\defchic}
$$
\left({R\over\ell_f}\right)_{col}=\chi_{c}\simeq0.569 \, . \eqno (\new)
$$
Note that $r_i\gg\ell_f$ is the physically relevant case for string dynamics in condensed
matter contexts (recall that the dynamics in that case is always friction-dominated). Also
note that because of friction, all loops will rapidly become  (almost) circular.

After numerically comparing the averaged and microscopic evolution equations, we find that the
simplest possibility (shown in fig.~3.1),
\eqnam{\fkans}
$$
k=\cases{1,&${R\over\ell_f}>\chi$ \cr {1\over\sqrt2}{R\over\ell_f},
&${R\over\ell_f}<\chi$ \cr} \, , \eqno (\new)
$$
provides excellent agreement for thse circular loops (see figures 3.2-3). In particular, this
turns out to be significantly better than assuming smoother (and slower) transitions between the
two regimes. As can be readily seen, this ansatz provides a very good fit, considering
the lack of parameters available.

\midinsert
\vbox{\centerline{
\epsfxsize=.7\hsize\epsfbox{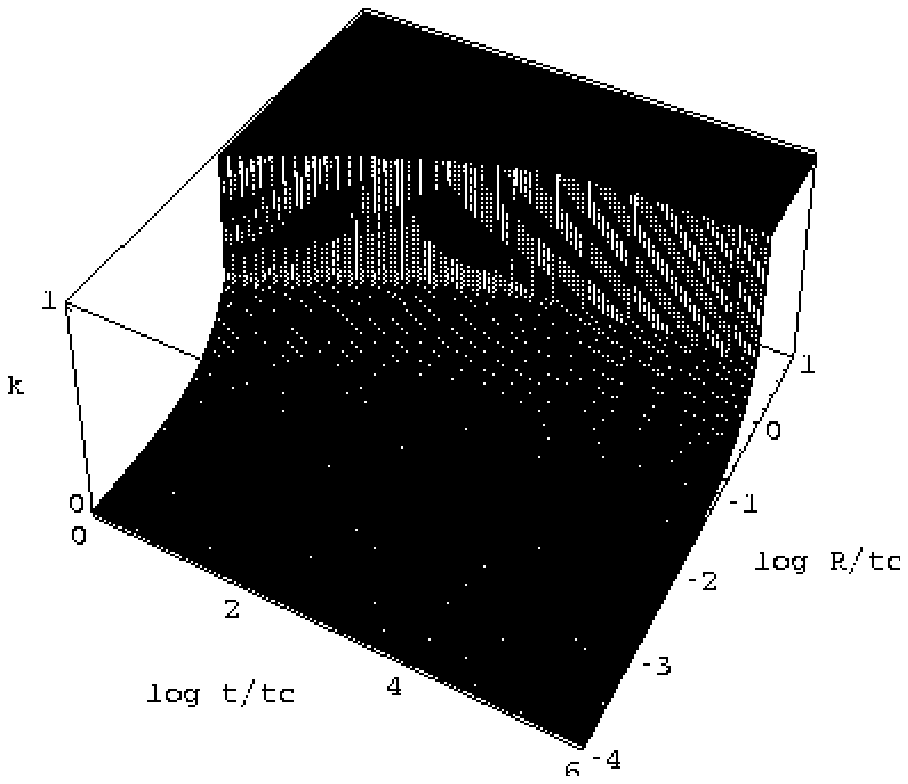}}
\vskip.5in
{\baselineskip 10pt\noindent\narrower\rm\hbox{\eightbf
Figure 3.4}:\quad\eightrm
{The ansatz for parameter $k$ for gauge GUT string loops,
as a function of loop radius and time after the formation of the string network. (see Section
5).}\smallskip}}
\endinsert

In passing, it is worth pointing out that one can also easily calculate the loop
lifetime\refto{gs}. In the relativistic regime, the ${\bar R}$ evolution equation can be
written
\eqnam{\avevrel}
$$
{d{\bar R}\over dt}=-{{\bar R}\over2\ell_f} \, , \eqno (\new)
$$
so we can immediately estimate that the loop will disappear in a time $t_{dec}\sim2\ell_f$
after its first collapse.

The case of the circular loop (\ccl) in the expanding universe is analogous, with the constant
friction length being replaced by a time-dependent damping length (\daml); also the invariant
loop radius is now $R=ar/\sqrt{1-{\dot r}^2}$.
Hence the microscopic evolution equations now take the form
\eqnam{\smallru}
$$
{\ddot r}+(1-{\dot r}^2)\left(a{{\dot r}\over\ell_d}+{1\over r}\right)=0 \, , \eqno (\new)
$$
\eqnam{\unavvu}
$$
{dv\over dt}=(1-v^2)\left({1\over r}-{v\over\ell_d}\right) \, , \eqno (\new)
$$
$$
{dR\over dt}=HR-v^2{R\over\ell_d} \, . \eqno (\new)
$$

\midinsert
\vbox{\centerline{
\epsfxsize=.7\hsize\epsfbox{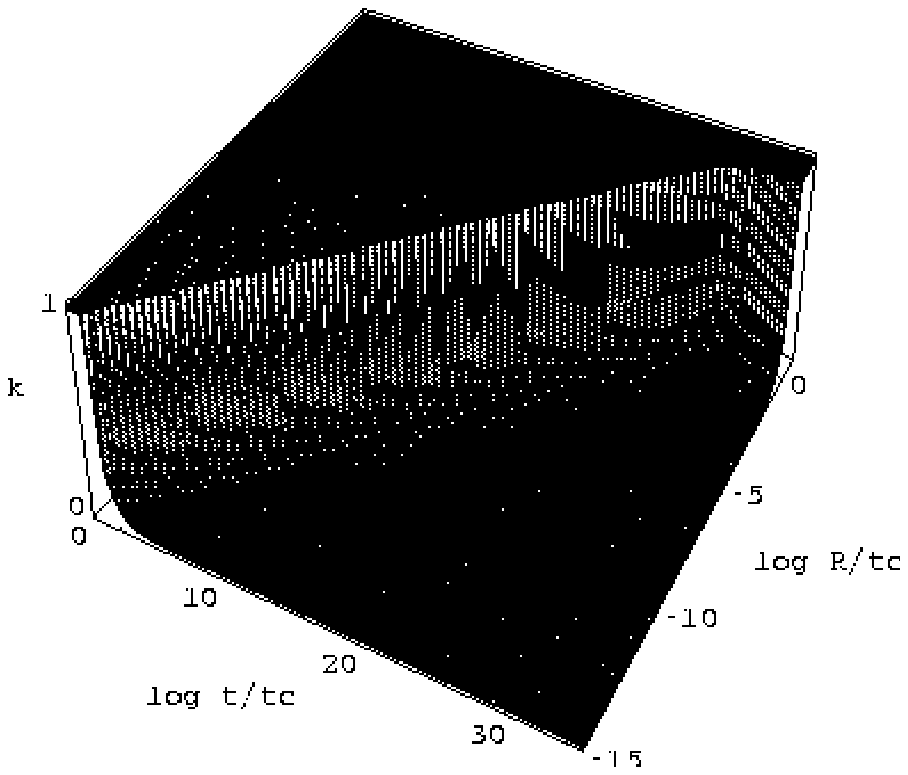}}
\vskip.5in
{\baselineskip 10pt\noindent\narrower\rm\hbox{\eightbf
Figure 3.5}:\quad\eightrm
{The ansatz for parameter $k$ for gauge electroweak string
loops (compare with figure 3.4). Note that these units $t_{eq}\sim22.1$,
friction ceases to dominate the dynamics at
$t_\ast\sim25.8$ and the present time is $t_o\sim28.5$.}\smallskip}}
\endinsert

\noindent Neglecting the gravitational radiation term (whose form has been established
elsewhere), the averaged evolution equations are
\eqnam{\avevu}
$$
{d{\bar R}\over dt}=(1-2{\bar v}^2)H{\bar R}-{\bar v}^2{{\bar R}\over\ell_f} \, , \qquad 
{d{\bar v}\over dt}=(1-{\bar v}^2)\left({k({\bar R})\over {\bar R}}-{{\bar
v}\over\ell_d}\right)
\, . \eqno (\new)
$$

Corresponding to this change, we simply modify $\ell_f$ to $\ell_d$ in our ansatz for $k$, which
becomes (\kans) (see figures 3.4-5). However, the numerical value of the ratio
of $R$ and $\ell_d$ when the loop first collapses, $\chi$, is now slightly
smaller:
\eqnam{\defchiu}
$$
\chi=\cases{0.431 \ ,& Radiation \cr 0.380 \ ,
& Matter} \, . \eqno (\new)
$$
Numerically we find that a slightly larger value, $\chi\sim0.5$ provides the
best fit in both cases. It is also interesting to note that while the
radiation-era evolution is rather insensitive to the value of $\chi$ (in the
range $\chi\sim0.3-0.6$ say), the matter-era evolution exhibits a stronger dependence.

\midinsert
\vbox{\centerline{%
\hskip4em\epsfxsize=.6\hsize\epsfbox{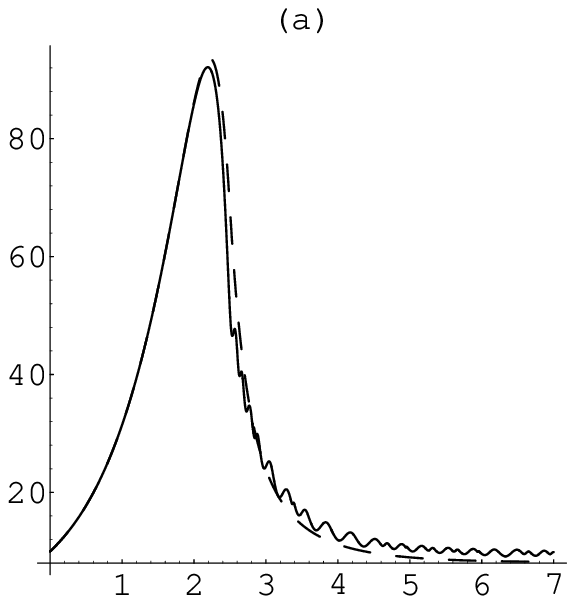}\hskip-10em\epsfxsize=.6\hsize\epsfbox{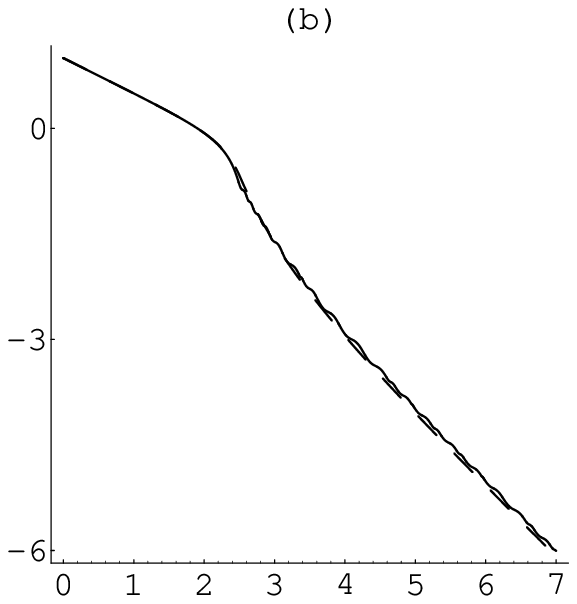}}
\vskip-.75in}
\endinsert
\midinsert
\vbox{\centerline{%
\hskip4em\epsfxsize=.6\hsize\epsfbox{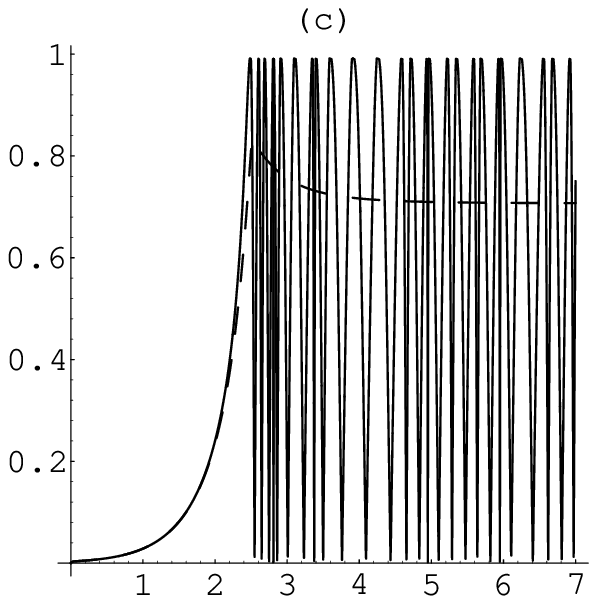}\hskip-10em\epsfxsize=.6\hsize\epsfbox{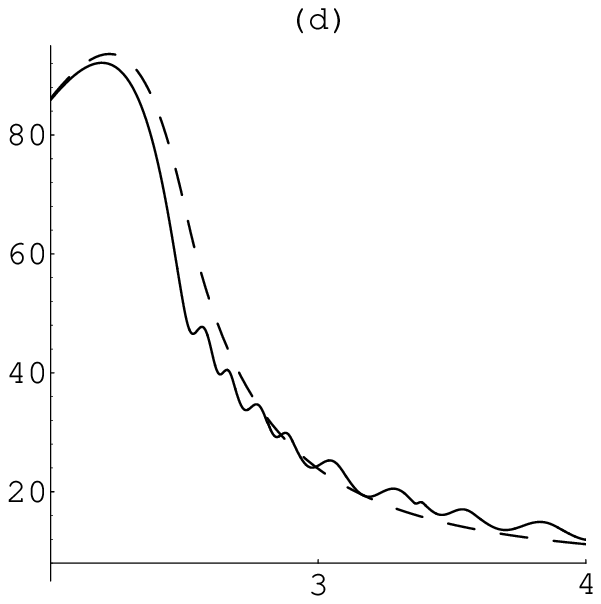}}
\vskip-.75in
{\baselineskip 10pt\noindent\narrower\rm\hbox{\eightbf
Figure 3.6}:\quad\eightrm
{Comparing the `microscopic' (solid) and `averaged' (dashed)
evolution equations for the physical radius (a), the radius relative to the horizon
(b), and velocity (c) of a circular gauge GUT loop formed at $t=t_c$ with radius $R=10t_c$; (d)
is a close-up of (a). Time is in orders of magnitude from the moment of
loop formation. Radiative mechanisms are not included.}\smallskip}}
\endinsert

Figures 3.6-7 depict the evolution of a loop with initial radius $R_i=10
t_i$ in the radiation ($t_i=t_c$) and matter ($t_i=t_{eq}$) eras---see
section 4. Again, the loops are initially overcritically damped, the
approximate velocity being
\eqnam{\fovdu}
$$
v\sim {\ell_d\over R} \, . \eqno (\new)
$$
The effect of damping is essentially twofold. Firstly, it delays the moment
when the loop first collapses. While its velocity is non-relativistic, there
is no loss of length through velocity redshift (see (\avevu)), and so the
physical loop radius can grow to a size much larger than the initial radius.
As it picks up speed, however, it starts losing more energy. As we
mentioned above, the first collapse still occurs for $R\sim\ell_d$; in the
relativistic regime, the loop loses energy at each oscillation as before.

Finally, when friction has switched off (and the period of oscillation is much
shorter than the expansion rate) the loop starts to oscillate with constant
physical amplitude (see (\avevu)). Notice that due to the
additional effect of friction this energy loss is much larger in the radiation
era. In this case the `final' physical radius is almost equal to the initial
radius, whereas in the matter era it can be more than one order of magnitude larger. 
When this final stage is reached,
gravitational radiation or other preferred decay channels cause the loop to shrink further.

\midinsert
\vbox{\centerline{%
\hskip4em\epsfxsize=.6\hsize\epsfbox{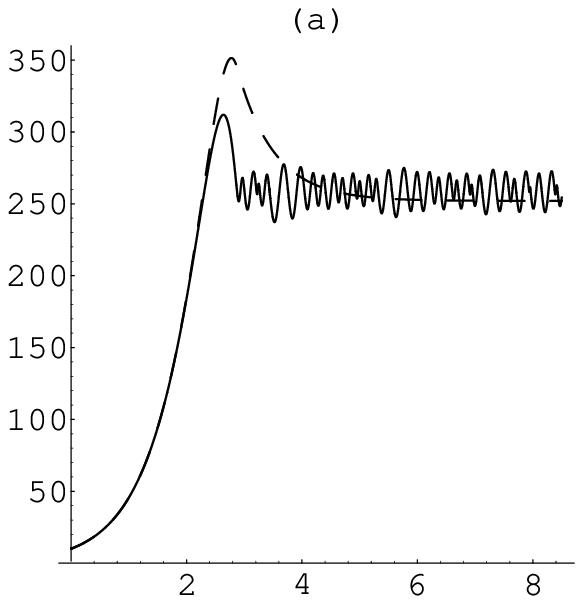}\hskip-10em\epsfxsize=.6\hsize\epsfbox{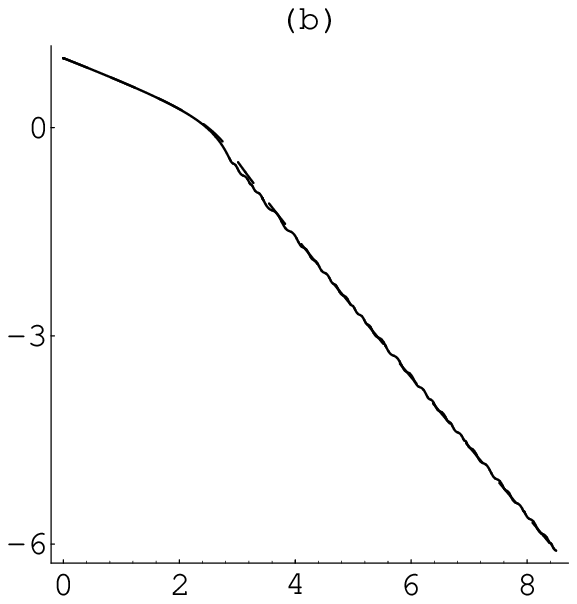}}
\vskip-.75in
{\baselineskip 10pt\noindent\narrower\rm\hbox{\eightbf
Figure 3.7}:\quad\eightrm
{Comparing the
`microscopic' (solid) and `averaged' (dashed) evolution equations for the
physical radius (a) and the radius relative to the horizon (b) of a circular
gauge GUT loop formed at $t=t_{eq}$ with radius $R=10t_{eq}$. Time is in
orders of magnitude from the moment of loop formation. Radiative mechanisms
are not included.}\smallskip}}
\endinsert

Having thus established, in simple but physically relevant cases, the validity of our
generalized `one-scale model', and in particular of the ansatz for $k$, we now proceed to
apply it to the study of cosmic string evolution; the study of string
evolution in condensed matter contexts is left for a companion
paper\refto{ms2}.

\sectbegin{4}{General scaling results}
\bigskip
\nobreak

\subbegin{A. Introduction}
\bigskip

\noindent In the early universe the friction lengthscale increases with time, so friction
will only be important at early times\refto{gs}; however, the meaning of `early' is, as we
will see, model-dependent. Again let $T_c$ be the temperature of the string-forming phase
transition; the corresponding time of formation is
\eqnam{\tf}
$$
t_c={1\over f}{m_{Pl}\over T^2_c} \, , \eqno (\new)
$$
where $f$ is given by
\eqnam{\deff}
$$
f=4\pi\left({\pi{\cal N}\over45}\right)^{1/2} \, , \eqno (\new)
$$
and ${\cal N}$ is the number of effectively massless degrees of freedom in the model (e.g.,
${\cal N}=106.75$ for a minimal GUT model, but it can be as high as $10^4$ for particular
extensions of it). Then in the case of a gauge symmetry breaking the friction lengthscale can be
written
\eqnam{\lfgauge} 
$$
\ell_{\rm f}=\cases{{1\over\theta}{t^{3/2}\over t_c^{1/2}},& Radiation era
\cr \left({3\over4}\right)^{3/2}{1\over\theta}{t^2\over \left(t_ct_{eq}\right)^{1/2}},&Matter era \cr}
\, , \eqno (\new)
$$
and for the case of a global symmetry
\eqnam{\lfglobal} 
$$
\ell_{\rm f}=\cases{{1\over4\theta}{t^{3/2}\over
t_c^{1/2}}\ln{\left({L\over\delta}\right)}\left[\ln{\left({6\over\lambda}{t_c\over
t}\right)}\right]^2,& Radiation era \cr \left({3\over4}\right)^{3/2}{1\over4\theta}{t^2\over
\left(t_ct_{eq}\right)^{1/2}}\ln{\left({L\over\delta}\right)}\left[\ln{\left({8\over\lambda}{t_ct_{eq}^{1/3}\over
t^{4/3}}\right)}\right]^2,&Matter era \cr} \, . \eqno (\new)
$$
The constant $\theta$ is a measure of the importance of the friction term in the evolution equations;
its value is
\eqnam{\valth} 
$$
\theta={\beta\over\sqrt{f}}\left(t_c\over t_{Pl}\right)^{1/2} \, . \eqno (\new)
$$
Note that as we mentioned previously, $\beta$ is not exactly the same in the gauge and global cases.
Also, (\lfgauge) is valid almost immediately after $t_c$ (so we will begin studying the
evolution of the network at $t_i\approx t_c$), but (\lfglobal) is only valid for
$t/t_c>6/\lambda$. The string energy per unit length can again be written
\eqnam{\mueu} 
$$
\mu=\cases{T_c^2,& Gauge case
\cr T_c^2\ln{\left({L\over\delta}\right)},&Global case \cr} \, . \eqno (\new)
$$

Defining $t_{\ast}$ as the
time at which the two damping terms in (\spc) and (\tc) have equal magnitude we find
\eqnam{\tstarr}
$$
{t_{\ast}\over t_c}=\cases{\theta^2,&Gauge case \cr
16\theta^2\left(\ln{{L\over\delta}}\right)^{-2}\left[\ln{\left({6\over\lambda}{t_c\over
t_{\star}}\right)}\right]^{-4},&Global case \cr} \, , \eqno (\new)
$$
provided this is still in the
radiation era; otherwise, in the matter era we obtain 
\eqnam{\tstarm}
$$
{t_{\ast}\over t_c}=\cases{\left({4\over3}\right)^{1/2}\theta\left({t_{eq}\over
t_c}\right)^{1/2},&Gauge case \cr
4\left({4\over3}\right)^{1/2}\theta\left({t_{eq}\over
t_c}\right)^{1/2}\left(\ln{{L\over\delta}}\right)^{-1}\left[\ln{\left({8\over\lambda}\left({t_{eq}\over
t_c}\right)^{1/3}\left({t_c\over t_{\star}}\right)^{4/3}\right)}\right]^{-2},&Global case \cr} \, .
\eqno (\new) $$
String dynamics is friction-dominated from $t_c$ until
$t_{\ast}$, after which motion will become relativistic or `free'. A simple heuristic argument
(see, for example\refto{hk}) due to Kibble suggests that in the damped phase the correlation
length will scale as $L\propto t^{5/4}$; it is worth noting that this argument is rather similar to
the ones used in condensed-matter contexts.

We now address the problem of initial conditions. Starting with the correlation length, by
causality this must obviously be smaller than the horizon; but on the other hand, it must be
greater than the friction lengthscale since friction is initially dominating the dynamics. We will
therefore assume (following\refto{hk}) that
\eqnam{\gic}
$$
\ell_{{\rm f}i}<L_i<t_i \, . \eqno (\new)
$$
For simplicity, we now concentrate on the gauge case, and define ${\tilde L}\equiv L/t_c$. Then
(\gic) can be written
\eqnam{\lic}
$$
\theta^{-1}<{\tilde L}_i<1 \, ; \eqno (\new)
$$
 these two extreme limits could correspond
to a rapid second-order phase transition ($\tilde L_i\sim \theta^{-1}$) or a slow first-order
transition ($\tilde L_i\sim1$). Now, the parameter $\beta$ in (\expb) can be written
\eqnam{\newb}
$$
\beta={2\zeta (3)\over\pi^2}{\cal N}\omega \, , \eqno (\new)
$$
(where $0\le\omega<1$) and using  the definitions of $\theta$ and $f$, one
finds that the initial ratio of the string and background densities obeys
\eqnam{\densi}
$$
{32\pi\over3}G\mu\le\left({\rho_{\infty}\over\rho_{b}}\right)_i\le{60\zeta (3)\over \pi^4}\omega
~~~(\,\le0.75\,)\, . \eqno (\new)
$$

In fact, by analyzing the evolution equations (\evl,\evv) one can show that for all physically
reasonable values of ${\tilde c}$ and $k$ these bounds hold for all subsequent times. Consequently,
cosmic strings can never dominate the universe. It can analogously be shown that this is also
true in the case of a global symmetry, although the required algebra is slightly more
complicated. Note that this is an entirely different (and independent) argument to Albrecht \&
Turok's\refto{albrecht} statistical physics argument---in a sense, it is a `thermodynamical'
argument.

As for the initial velocity, it can be estimated by a very simple (and rough) argument. The friction
lengthscale can be approximately interpreted as the distance that a piece of string can travel
before it is stopped by the friction force. Then by comparing the energy contained in strings in a
volume $L^3$ and the work done by the friction force in stopping them we obtain
\eqnam{\idev}
$$
\mu L_i\sim {\mu v_i\over\ell_{{\rm f}_i}}L_i^2 \, , \eqno (\new)
$$
and substitution of $\ell_{\rm f}$ yields
\eqnam{\vest}
$$
v_i{\left({L\over t}\right)}_i\sim \left(f{t_{Pl}\over t_c}\right)^{1/2} \, . \eqno (\new)
$$
As expected, highly-curved strings will have high velocities and conversely strings in low-density
networks will have small velocities.

\subbegin{B. Long-string scaling laws}
\bigskip

\noindent Analysis of the evolution equations for the correlation length and velocity of the
long string network (\evl,\evv) reveals the existence of three types of scaling regimes,
which we now describe in detail. Two of these regimes are transient, occurring in the
friction-dominated epoch. In this situation strings should have very little
small-scale structure, and we should have
$k=1$. In analogy with our discussion for string loops in section 3, we
should also find that the string velocity in the radiation era is
\eqnam{\vapprfr}
$$
v\propto {\ell_d\over L} \, , \eqno (\new)
$$
which is in fact the case (see below). The third is the well-known linear scaling regime.

Recall that $t_c$ (defined in (\tf)) denotes the time of string
formation; on the other hand, $t_s$ will denote the time at which the relevant period of
evolution starts.
\medskip
\subhead{Stretching regime}\noindent This is a transient regime that occurs in the beginning of
the friction-dominated phase, provided the initial string density is sufficiently low. With
$t_s=t_c$ we get
\eqnam{\strl}
$$
L=L_s\left({t\over t_s}\right)^{1/2} \, , \eqno (\new)
$$
\eqnam{\strv}
$$
v=\cases{{t\over\theta L_s} ,&Gauge case \cr {t\over\theta L_s}
\left[\ln\left({\lambda\over6}{t\over
t_s}\right)\right]^2\ln{L\over\delta},&Global case \cr} \, , \eqno (\new)
$$

The reason why this regime arises is physically obvious. If we start with a small string
density, the correlation length will be close to the horizon, and much larger than the damping
length. Hence long strings are fairly straight and have very little small-scale structure.
This is therefore analogous to the situation in condensed matter---the strings have very small
velocities and are `frozen', being conformally stretched by the expansion. 

Mathematically, the
$L\propto t^{1/2}$ law comes immediately from the fact that the friction, loop production and
redshift terms in the evolution equation for $L$ are all velocity-dependent, and can
therefore be neglected when compared to expansion. Note, however, that unlike the condensed
matter case there is no additional logarithmic correction to the $L\propto t^{1/2}$ law in the
case of a global string network.
Finally, and although this is not cosmologically relevant, it should be pointed out that the
corresponding regime in the matter era would be 
\eqnam{\stretchmat}
$$
L\propto t^{2/3} \,, \qquad v\propto t^{4/3} \, . \eqno (\new)
$$

\medskip\subhead {Kibble regime}\noindent This is a transient regime that also occurs in the
friction-dominated epoch (following the stretching regime when such a regime exists). In this
case the scaling laws are
\eqnam{\kibl}
$$
{L\over t_c}=\cases{\left[{2(1+{\tilde c})\over3\theta}\right]^{1/2}\left({t\over
t_c}\right)^{5/4} ,&Gauge case \cr \left[{2(1+{\tilde
c})\over3\theta}\right]^{1/2}\left({t\over t_c}\right)^{5/4}\ln{\left({\lambda\over6}{t\over
t_c}\right)}\left(\ln{L\over\delta}\right)^{1/2},&Global case \cr} \, , \eqno (\new)
$$
\eqnam{\kibv}
$$
v=\cases{\left[{3\over2\theta (1+{\tilde c})}\right]^{1/2}\left({t\over
t_c}\right)^{1/4} ,&Gauge case \cr \left[{3\over2\theta (1+{\tilde
c})}\right]^{1/2}\left({t\over t_c}\right)^{1/4}\ln{\left({\lambda\over6}{t\over
t_c}\right)}\left(\ln{L\over\delta}\right)^{1/2},&Global case \cr} \, , \eqno (\new)
$$

This is a high-density regime, arising when the correlation length $L$ is close to the
friction length---either because it started that way or because it becomes so during a period
of $L\propto t^{1/2}$ evolution (recall that in the radiation era the friction length grows as
$\ell_f\propto t^{3/2}$).

Although friction still dominates the dynamics, the higher string curvature (and consequently
higher velocity) means that the network is now chopping off a considerable amount of energy
into loops---in fact, proportionally more than in the final, linear regime. However, note that
there is still no small-scale structure (ie, we still have $k\approx1$). Therefore this shows that
small-scale structure is not the only determining factor for loop production---the long-string
density (ie, $L$) is just as important. This is the reason why we chose the form of our loop
production ansatz (\inls), explicitly separating the effects of large (through
$L$) and small-scale structure (through the parameter $\alpha$).

The $L\propto t^{5/4}$ scaling law was previously suggested by Kibble. However, note that the
above relations only hold in the radiation era. In the corresponding regime in the matter era
for very light strings, we have the scaling laws
\eqnam{\kibrmat}
$$
L\propto t^{3/2} \,, \qquad v\propto t^{1/2} \, . \eqno (\new)
$$
\medskip\subhead {Linear regime}\noindent This is the well-known
`final' scaling regime---it is always the endpoint of cosmological
string evolution, arising when the friction lengthscale becomes subdominant
with respect to the Hubble length. Assuming that
$k$ (as well as ${\tilde c}$) are constant in each regime, we find
the following scaling laws
\eqnam{\linl}
$$
L=\cases{\left[k_r(k_r+{\tilde c_r})\right]^{1/2}t,&Radiation era \cr  \left[9k_m(k_m+{\tilde
c_m})\over8\right]^{1/2}t,&Matter era \cr} \, , \eqno (\new)
$$
\eqnam{\linv}
$$
v=\cases{\left[k_r\over (k_r+{\tilde c_r})\right]^{1/2} ,&Radiation era \cr 
\left[k_m\over2(k_m+{\tilde c_m})\right]^{1/2},&Matter era \cr} \, . \eqno (\new)
$$

Now, the simplest way to proceed is to look for the values of ${\tilde
c}$ and $k$ that match the simulations: \eqnam{\rad}  $$
\eqalign{\tilde c}_r\approx0.24 \, , \qquad k_r\approx0.18 \, , \eqno (\new)
$$
\eqnam{\mat} 
$$
\eqalign{\tilde c}_m\approx0.17 \, , \qquad k_m\approx0.49 \, . \eqno (\new)
$$
Hence, according to our interpretation of this model, it predicts a larger loop production rate
and more small-scale structure in the radiation era (recall that more small-scale structure
corresponds to a smaller
$k$)---which is exactly what is seen in numerical simulations\refto{bb}. This shows that our
interpretation of
$k$ as being related to the presence of small-scale structure is at least qualitatively correct.
Notice that the scaling parameters are much less sensitive to variations in
${\tilde c}$ than those of previous analytic models; this will be relevant below.

Furthermore, we can get some feeling for the validity of our ansatz (\kans)
for $k$ by
finding out which values we obtain with the known values
of the ratios $\zeta=L/t$ in the radiation and matter eras\refto{bb}. In the first case we have
$\zeta_r\sim0.28 \, , k_r\sim0.19$ in excellent agreement with (\rad). In the matter
era the scaling value of
$\zeta_m\sim0.535 $ also gives good agreement with the simulations, since
$k_m=0.50$ with
$\chi=1$ (the dashed line ansatz in fig.~3.1). Note, however, that matter era
correspondence is upset if we use the $\chi=0.57$ ansatz modified to match
collapsing circular loops.  Nevertheless, these results indicate that the ansatz (\kans),
originally justified for string loops, is probably extensible to the long string network,
although the point $\chi$ where the transition between the constant and linear regimes takes
place should be studied in greater detail. If this is indeed the case, then it appears that this
simple model goes a long way towards solving the `matching' problem at the matter-radiation
transition---previously, a significant failing of the `one-scale' model.

Of course, these results are a manifestation of
the fact that we require additional degrees of freedom to incorporate small-scale structure
satisfactorily (see, for example, ref.\refto{ack}), but perhaps there are fewer required than
at first anticipated.

\subbegin{C. Loop scaling laws}
\bigskip

\noindent We already know how to determine the (relative) loop density at all times; we simply
have to evaluate
\eqnam{\ra2tio}
$$
{\varrho(t)}=g{\tilde
c}L^2(t)\int_{t_c}^{t}{v_\infty (t')\over L^4(t')}{\ell(t,t')\over\alpha (t')}dt'
\, ,
\eqno (\new)
$$
where $t_c$ is the network formation time and $\ell(t,t')$ is the length at time $t$
of loops produced at time $t'$. If necessary, we can also analyze the
distributions of loop lengths, etc.---although this will not be done in this
paper. There is still, however, one point to be discussed---we must propose an
ansatz for the parameter
$\alpha$, which should be related to the presence of string small-scale structure.

We know on physical grounds that $\alpha\sim1$ in the friction
dominated epoch, since strings are non-relativistic and any wiggles are
quickly erased; this conclusion is also supported qualitatively by observing the actual evolution
of networks in quenched liquid crystals\refto{cdty}.
On the other hand, all studies of radiative backreaction lead us to expect $\alpha$ to be a
constant (significantly less than unity) in the linear scaling regime. For example, in the case
of a gauge GUT string network numerical simulations\refto{bb} show that
$\alpha_{GUT}<5\times10^{-3}$ (note that our definition of
$\alpha$ differs from that in the numerical simulation papers---where it is defined as
$\ell/t$). We will therefore use the simple ansatz
\eqnam{\alfans}
$$
\alpha(t)={1+\alpha_{sc}{t\over t_\ast}\over1+{t\over t_\ast}}\, ,
\eqno (\new)
$$
where $\alpha_{sc}$ is the constant scaling value. Note that this is such that
in the transition between the damped and free regimes the physical loop
length at formation ($\ell=\alpha L$) is constant. In this simple way we
can `phenomenologically' account for the build-up of small-scale structure as
seen in the numerical simulations\refto{bb}. In particular, for the case of
GUT strings, this build-up takes about 4 orders of magnitude in time, in
agreement with a result obtained by Allen
\& Caldwell and by Austin\refto{aca} by estimates of the evolution of the
linear kink density in a `one-scale' model context.

\midinsert
\vbox{\centerline{
\epsfxsize=.7\hsize\epsfbox{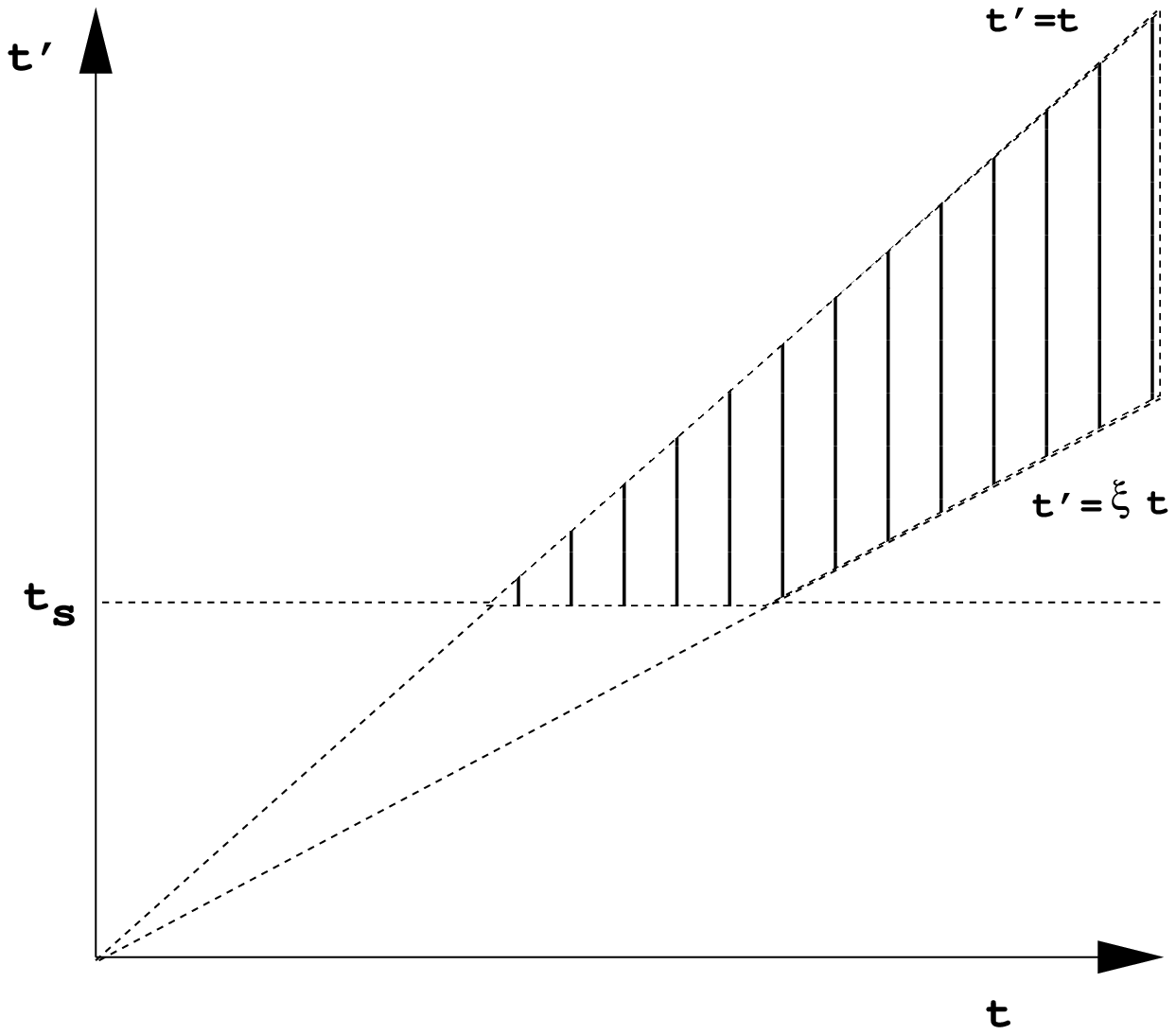}}
\vskip.5in
{\baselineskip 10pt\noindent\narrower\rm\hbox{\eightbf
Figure 4.1}:\quad\eightrm
{The interval in $t'$ giving a non-zero loop
length contribution to the loop density integral (\ra2tio) for different times
$t$.}\smallskip}}
\endinsert

Although (\ra2tio) cannot be evaluated analytically in general due to the complicated behavior
of the integrand (and in particular of the $\ell(t,t')$ factor), it is possible (under some
simplifying assumptions) to obtain an analytic solution in the linear
regime. For simplicity we will neglect the loops
existing at the start of this regime,
$t=t_s$.
Since loops are much smaller than the horizon, we can approximately neglect the effect of
expansion; furthermore, we can also assume that the loop velocity is close
to the terminal velocity, $v_\ell\sim1/\sqrt{2}$, so that we approximate
$\Gamma'v^6\sim\Gamma$. These two simplifying assumptions mean that our
calculation will be a slight underestimate of the true loop density. The loop decay function
in this case is the following
\eqnam{\loopdecay}
$$
\ell(t,t')=\cases{\ell'-\Gamma G\mu (t-t'),&$t'\le t\le t'+{\ell'\over\Gamma G\mu}$ \cr 
0,& Otherwise \cr} \, , \eqno (\new)
$$
with the initial loop length being
\eqnam{\initialloopdecay}
$$
\ell'=\alpha \zeta t' \, , \eqno (\new)
$$
and we used $\zeta=L/t$. Note that the coefficient $\Gamma G\mu$ will be altered if the 
preferred loop decay channel is not gravitational radiation, but Goldstone bosons or 
electromagnetic radiation. With these assumptions, 
(\ra2tio) reduces to
\eqnam{\ra2tio}
$$
\varrho(t)={g{\tilde
c}v_\infty\over\alpha_{sc} \zeta^2}t^2\int_{f(t)}^{t}{\ell(t,t')\over t'^4}dt'
\, ,
\eqno (\new)
$$
where
\eqnam{\ffunctionf}
$$
f(t)=\cases{t_s,&$t\le {t_s\over\xi}$ \cr 
\xi t,& $t\ge {t_s\over\xi}$ \cr} \, , \eqno (\new)
$$
and
\eqnam{\defcsi}
$$
{1\over\xi}=1+{\alpha \zeta\over\Gamma G\mu} \, . \eqno (\new)
$$

\midinsert
\vbox{\centerline{
\epsfxsize=.7\hsize\epsfbox{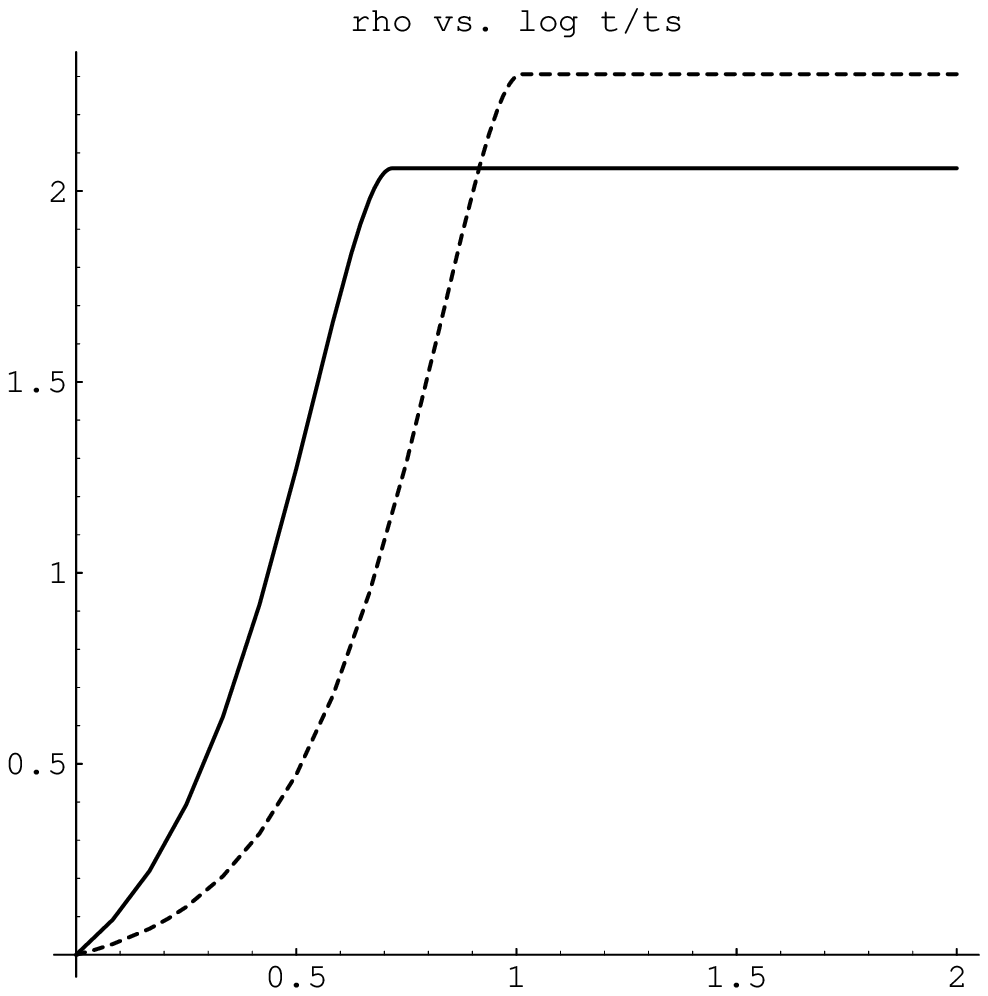}}
\vskip.5in
{\baselineskip 10pt\noindent\narrower\rm\hbox{\eightbf
Figure 4.2}:\quad\eightrm
{Analytical estimate of the evolution
of the ratio of the loop and long string energy densities in the linear
scaling regime, for
$\alpha=10^{-3}$ and $\Gamma G\mu=65\times 10^{-6}$, in the radiation (solid) and matter
(dashed) eras. Time is in orders of magnitude from the start of the regime.}\smallskip}}
\endinsert

The path of integration for all times is schematically represented in figure
4.1. The integral can be therefore conveniently split into two cases, the
transition point corresponding to the moment of the decay of the loops formed
at $t_s$. Then one finds the following solutions
\eqnam{\soldenr}
$$
{\rho_o(t)\over\rho_{\infty}(t)}={g{\tilde
c}v_\infty\over\alpha_{sc} \zeta^2}\times\cases{\left[{1\over2}(\zeta\alpha_{sc}+\Gamma
G\mu)\left({t^2\over t_s^2}-1\right)-{1\over3}\Gamma
G\mu\left({t^3\over t_s^3}-1\right)\right],&$t\le {t_s\over\xi}$
\cr 
\left[{1\over2}(\zeta\alpha_{sc}+\Gamma
G\mu)\left({1\over\xi^2}-1\right)-{1\over3}\Gamma
G\mu\left({1\over\xi^3}-1\right)\right],& $t\ge {t_s\over\xi}$ \cr} \, . \eqno (\new)
$$

\midinsert
\vbox{\centerline{
\epsfxsize=.7\hsize\epsfbox{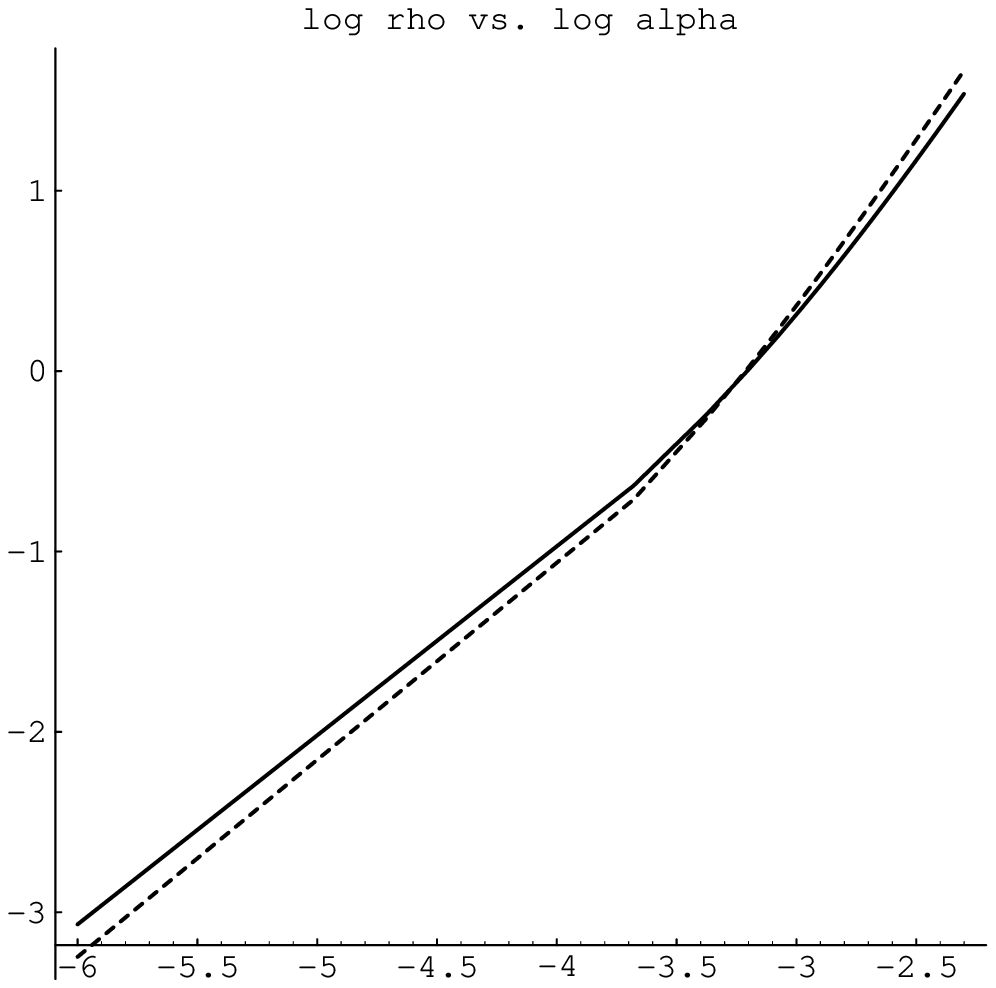}}
\vskip.5in
{\baselineskip 10pt\noindent\narrower\rm\hbox{\eightbf
Figure 4.3}:\quad\eightrm
{Expected ratios of the loop and long
string densities in the linear scaling regime, in the radiation (solid) and matter (dashed)
eras for gauge GUT strings, in terms of the small-scale structure parameter $\alpha$ and with
$\Gamma G\mu=65\times 10^{-6}$.}\smallskip}}
\endinsert

Again, this can be confirmed numerically (see section 5).
Figure 4.2 depicts a typical situation; after an initial build-up while no
loops have decayed, we reach a constant ratio. Note that (\soldenr) should be
testable against numerical simulations. With the values of ${\tilde c}$ and
$k$ quoted above and assuming that
$\alpha=10^{-3}$ and
$\Gamma G\mu=65\times 10^{-6}$ this model predicts the scaling density ratios to be
\eqnam{\solrat}
$$
{\rho_o(t)\over\rho_{\infty}(t)}=\cases{2.1,& Radiation era
\cr 
2.3,& Matter era \cr} \, . \eqno (\new)
$$

Notice that it is possible (at least in principle) to measure both $\alpha$ and the ratio of
the densities from numerical simulations ($\Gamma$ can also be estimated in this way).
Hence the prediction of this model for the density ratio in the scaling regime (\soldenr) and
even the approach to it can be tested numerically. Figure 4.3 shows the
predicted ratios for gauge GUT strings in the radiation and matter eras, at
fixed $\Gamma G\mu$ for all possible values of $\alpha$ (the upper limit is
inferred from numerical simulations, the lower limit assumed on physical
grounds).  Note that if loops are large enough, then  their density can be dominant (by a
factor of up to 50) and the ratio will be even higher in the matter era, converse to the
opposite limit for small loops. The reason is that in the matter era
loops live longer (see section 3), a factor compensated by the lower loop
production rate.

Finally, let us just briefly mention two other helpful quantities in characterizing the
evolution of a cosmic string network. The first is the fraction of the energy density in long
strings at a time $t$ which is converted into loops within the next Hubble time; this is given
by
\eqnam{\hubtf}
$$
f_H=g {{\tilde c}\over H} {v_\infty\over L}\, ,
\eqno (\new)
$$
and the limiting values for GUT strings in the radiation and matter eras are respectively
\eqnam{\solratf}
$$
f_H\sim\cases{0.81,&Radiation era
\cr 
0.18,&Matter era \cr} \, . \eqno (\new)
$$
The second is the number of loops chopped off by the long string network per Hubble volume per
Hubble time, given by
\eqnam{\hubtn}
$$
n=g {{\tilde c}\over\alpha} {v_\infty\over H^4L^4}\, ,
\eqno (\new)
$$
with asymptotic values
\eqnam{\solratn}
$$
n=\cases{2.2\times10^5,&Radiation era
\cr 
4.5\times10^3,&Matter era \cr} \, . \eqno (\new)
$$
These quantities should also be measurable in numerical simulations.

\midinsert 
\def\entry#1#2#3#4#5{\noalign{\hrule} &$#1$ && $#2$ && $#3$ &&
$#4$ && $#5$ & \cr\noalign{\hrule}}

$$\hbox{\vbox{\offinterlineskip
\def\strut{\hbox{\vrule height 10pt depth 6pt width 0pt}}
\hrule
\halign{
\strut\vrule # \tabskip 0.1in &
\hfil#\hfil &
\vrule$\,$\vrule #&
\hfil#\hfil &
\vrule # &
\hfil#\hfil &
\vrule # &
\hfil#\hfil &
\vrule # &
\hfil#\hfil &
\vrule #\tabskip 0.0in\cr\noalign{\hrule}
& {\bf Properties} \hskip 0.1in && Gauge EW && Global Ax. && Global GUT && Gauge GUT &
\cr\noalign{\hrule}\noalign{\hrule}
\entry{T_c/GeV}{10^2}{10^{10}}{10^{15}}{10^{16}}
\entry{G\mu}{10^{-34}}{10^{-18}}{10^{-8}}{10^{-6}}
\entry{\lambda}{Irrelevant}{0.2}{0.2}{Irrelevant}
\entry{\theta}{2\times 10^{15}}{1\times 10^8}{1\times 10^3}{29} 
\entry{t_c/t_{Pl}}{4\times 10^{32}}{3\times 10^{16}}{3\times 10^6}{3\times 10^4}
\entry{t_i/t_c}{1}{60}{60}{1}
\entry{t_{\star}/t_c}{7\times 10^{25}}{2\times 10^8}{351}{855}
\noalign{\hrule}}}}$$

\caption{Table 5.1}{Some relevant scales for the evolution of the four
cosmologically interesting string networks.}

\endinsert
\midinsert 
\def\entry#1#2#3#4#5{\noalign{\hrule} &$#1$ && $#2$ && $#3$ &&
$#4$ && $#5$ & \cr\noalign{\hrule}}

$$\hbox{\vbox{\offinterlineskip
\def\strut{\hbox{\vrule height 10pt depth 6pt width 0pt}}
\hrule
\halign{
\strut\vrule # \tabskip 0.1in &
\hfil#\hfil &
\vrule$\,$\vrule #&
\hfil#\hfil &
\vrule #&
\hfil#\hfil &
\vrule # &
\hfil#\hfil &
\vrule # &
\hfil#\hfil &
\vrule #\tabskip 0.0in\cr\noalign{\hrule}
& {\bf Case} \hskip 0.1in && Gauge EW && Global Ax. && Global GUT && Gauge GUT &
\cr\noalign{\hrule}\noalign{\hrule}
\entry{Solid}{10^{-1}}{0.9}{0.9}{0.9}
\entry{Dashed}{10^{-4}}{5\times 10^{-2}}{0.2}{0.2}
\entry{Dotted}{10^{-8}}{10^{-3}}{0.02}{0.05}
\noalign{\hrule}}}}$$

\caption{Table 5.2}{Initial conditions
on the ratio of the lengthscale to the horizon, $(L/t)_i$, to be used in the figures and
discussion below.}

\endinsert

\sectbegin{5}{The cosmologically relevant networks}
\bigskip
\nobreak

\subbegin{A. Introduction}
\bigskip

\noindent We can now discuss the four cosmologically interesting cases : gauge
electroweak and GUT, and global axionic and GUT strings. To summarize the
cases to be studied numerically, some relevant quantities in each case are
listed in table 5.1. The initial conditions taken for the correlation length
(expressed in terms of the ratio $L/t$) are summarized in table 5.2. Note that
the initial condition on velocity is not independent---see (\vest) in section
4.

There is another parameter that we must estimate, namely the typical loop size at
formation in the linear scaling regime, $\alpha_{sc}$. As we have seen in the above
discussion, this is crucial in determining the ratio of the energy densities in loops and long
strings. We will assume that $\alpha_{sc}\sim G\mu$.

\midinsert
\vbox{\centerline{%
\hskip4em\epsfxsize=.6\hsize\epsfbox{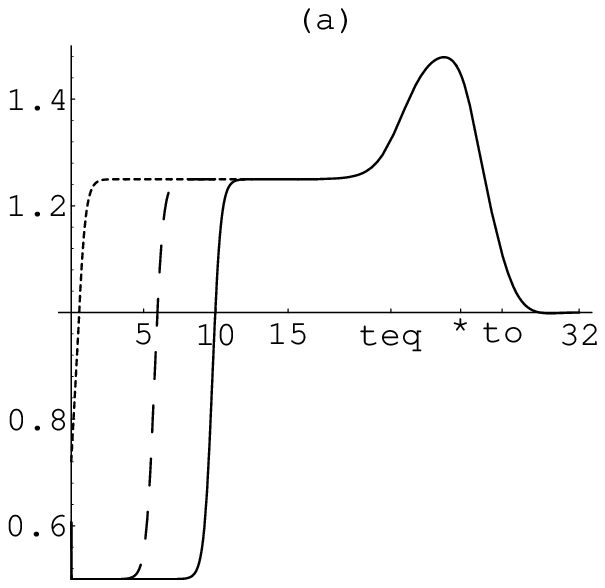}\hskip-10em\epsfxsize=.6\hsize\epsfbox{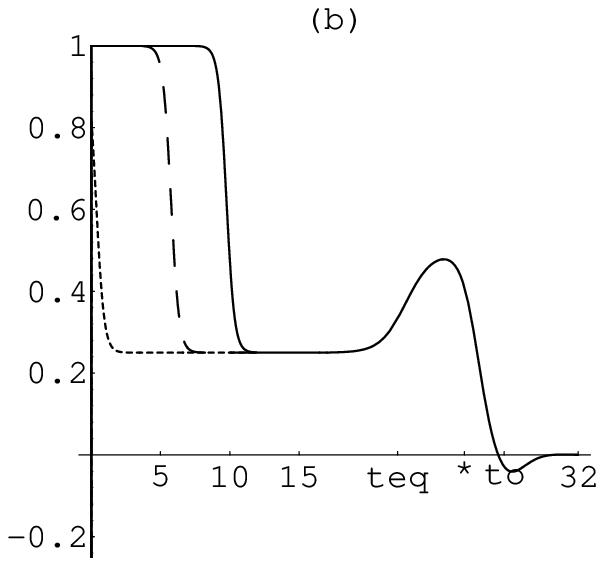}}
\vskip-.75in}
\endinsert
\midinsert
\vbox{\centerline{%
\hskip4em\epsfxsize=.6\hsize\epsfbox{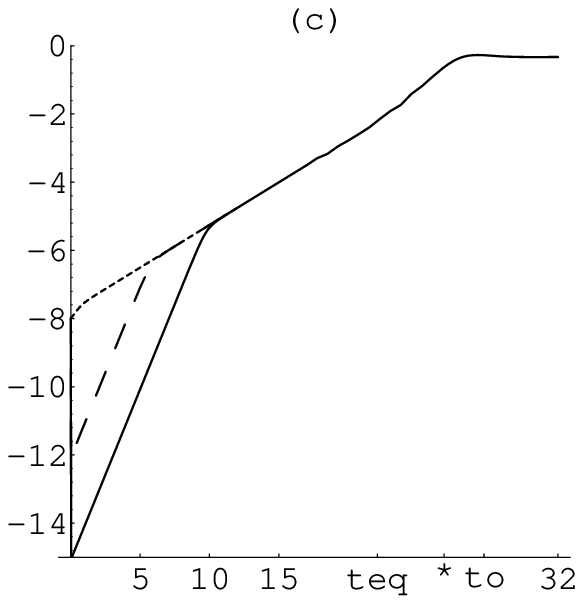}\hskip-10em\epsfxsize=.6\hsize\epsfbox{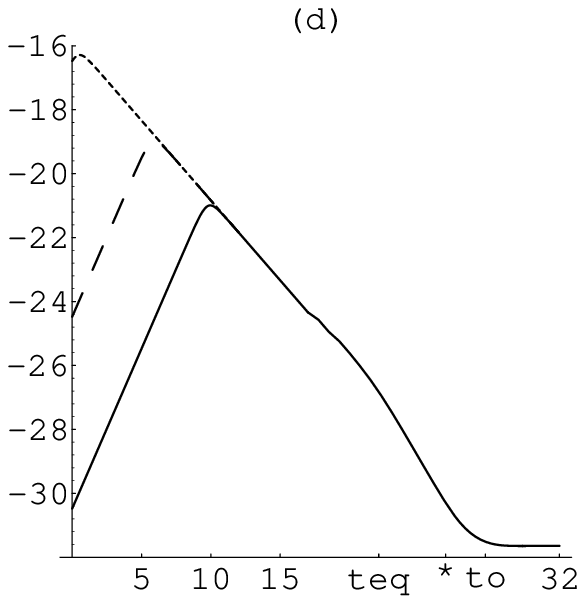}}
\vskip-.75in
{\baselineskip 10pt\noindent\narrower\rm\hbox{\eightbf
Figure 5.1}:\quad\eightrm
{The complete
evolution of an electroweak gauge long-string network. Plots successively
represent the exponents of the power-law dependence of $L$ (a) and $v$ (b),
$v$ itself (c), and the ratio of the long string and background densities (d).
The horizontal axis is labeled in orders of magnitude in time from the moment
of string formation, and initial conditions are specified in table
5.2.}\smallskip}}
\endinsert

\subbegin{B. Electroweak and axionic strings}
\bigskip

We start by noting that for electroweak and axionic strings we cannot meaningfully treat the
case of very high string density (that is, the case where $L$ is initially close to the
friction lengthscale). This is because in this case loop
reconnections onto the string network may have a significant effect. In all other situations the
effect of such reconnections can be neglected.

For gauge electroweak strings (see figures 5.1-2) the epoch of
friction-dominated dynamics ends well into the matter era. One of the most
important results on this paper is that the early-time evolution of the
network depends crucially on the initial string density. Note that among other
things this depends on the order of the phase transition at which the strings form. If this
density is low, strings have very small velocities and according to our previous discussion the
network starts evolving in the stretching regime, which can last up to ten orders of magnitude in
time. In this regime, the energy transfer into loops is negligible; this can be seen in 
figure 5.2(b), where initially less than one loop is
produced per Hubble time.

\midinsert
\vbox{\centerline{%
\hskip4em\epsfxsize=.6\hsize\epsfbox{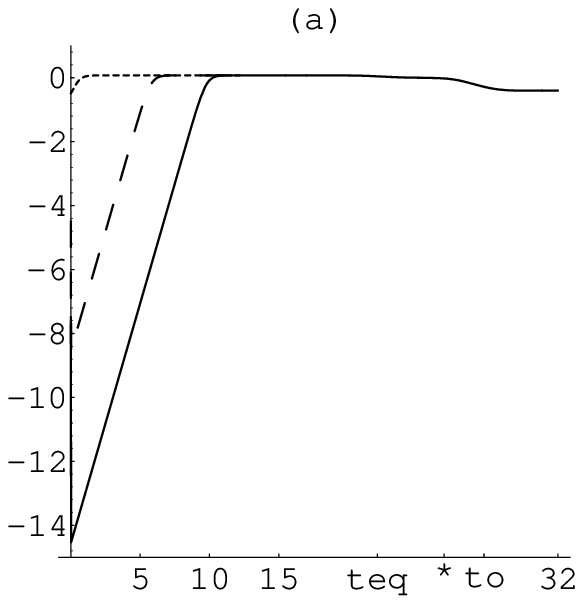}\hskip-10em\epsfxsize=.6\hsize\epsfbox{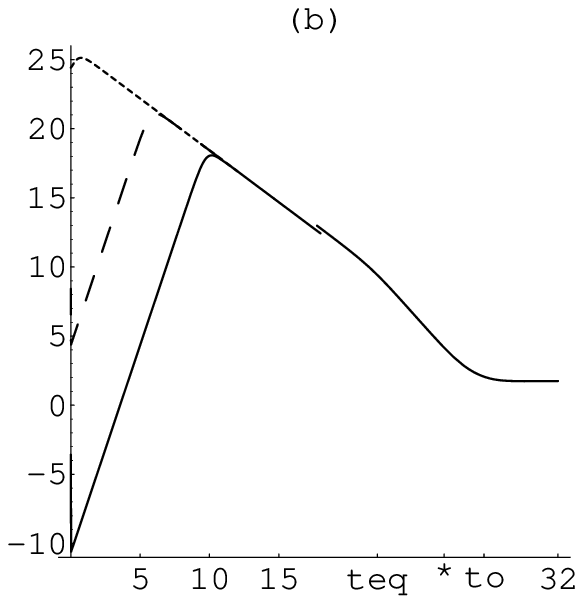}}
\vskip-.75in
{\baselineskip 10pt\noindent\narrower\rm\hbox{\eightbf
Figure 5.2}:\quad\eightrm
{Characteristics of
loop production by electroweak strings: log plots of the fraction of the
energy density going into loops per Hubble time (a) and of the number of loops
produced per Hubble volume per Hubble time (b). The horizontal axis is
labeled in orders of magnitude in time from the moment of string formation,
and initial conditions are specified in table 5.2; we have taken
$\alpha\sim10^{-31}$ and $\Gamma G\mu\sim10^{-30}$.}\smallskip}}
\endinsert

As the string velocity increases, energy losses to loop production become more
and more important and eventually the network `switches' rather quickly
to the Kibble regime. Here, the fraction of the energy density in long strings that is
transferred into loops per Hubble time is in fact larger than unity! Of course there is
nothing unphysical about this, since the string network also gains energy during the
time interval in question due to stretching---most of which is immediately converted into loops.
Note that in this regime, although the fraction of transferred energy is constant, the number of
loops produced per Hubble volume per Hubble time decreases: this is because the correlation
length is `catching up' with the horizon, so larger and larger loops (relative to the horizon )
are produced, given our assumption that during the friction-dominated epoch
$\alpha\approx1$.

In the opposite regime for high initial densities, the initial velocity is much larger, so loop
formation is important right from the start and the Kibble regime begins immediately. Note that
the first few orders of magnitude in time after the formation of an electroweak string network
are the relevant period for baryogenesis mechanisms; we therefore believe that
these results can shed some new light in this area.

As we approach $t_{eq}$, the network scaling switches again to a different regime, though now
rather more slowly. This would be the matter-era analogous of the Kibble regime ($L\propto
t^{3/2}$,
$v\propto t^{1/2}$), except that it is not particularly distinct given that the friction-dominated
dynamics ends and the network evolves towards the final linear scaling regime. As we
have already pointed out, the long-string network reaches the linear scaling
regime about today.  Nevertheless, it is still building up small-scale structure and,
if our ansatz is valid, it will keep on doing so for another 20 orders of
magnitude in time if gravitational radiation were the dominant decay mechanism (electromagnetic
losses would intervene first for superconducting strings).

\midinsert
\vbox{\centerline{%
\hskip4em\epsfxsize=.6\hsize\epsfbox{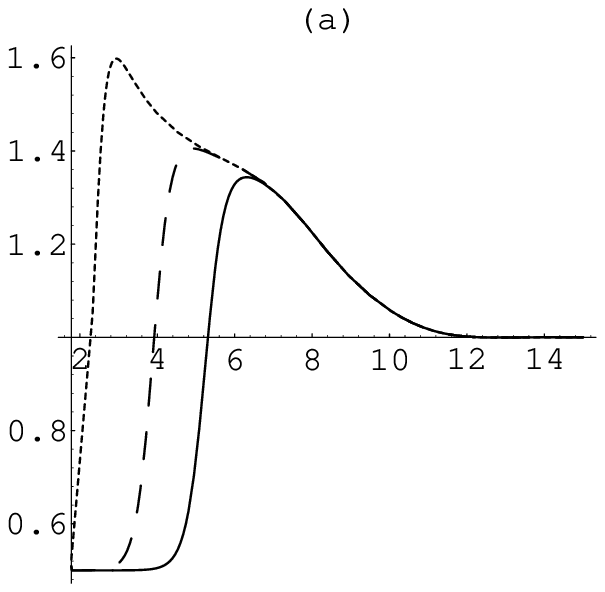}\hskip-10em\epsfxsize=.6\hsize\epsfbox{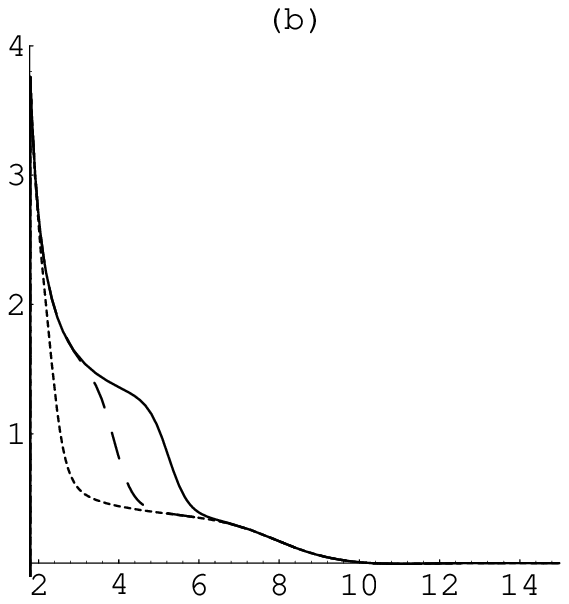}}
\vskip-.75in}
\endinsert
\midinsert
\vbox{\centerline{%
\hskip4em\epsfxsize=.6\hsize\epsfbox{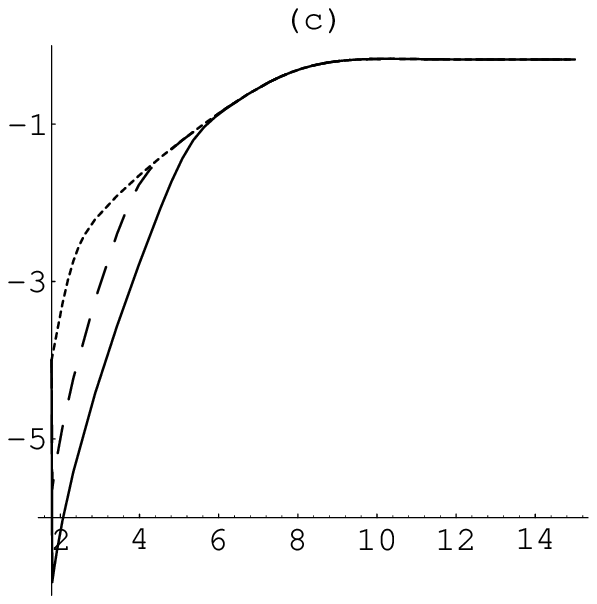}\hskip-10em\epsfxsize=.6\hsize\epsfbox{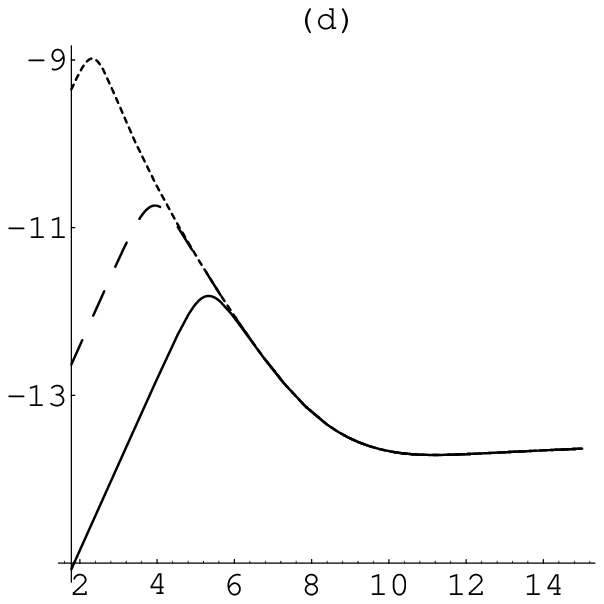}}
\vskip-.75in
{\baselineskip 10pt\noindent\narrower\rm\hbox{\eightbf
Figure 5.3}:\quad\eightrm
{The approach to
scaling  for an axionic long-string network. Plots successively represent the
exponents of the power-law dependence of $L$ and $v$, $v$ itself, and the
ratio of the long string and background densities. The horizontal axis is
labeled in orders of magnitude in time from the moment of string formation,
and (c) and (d) are log plots; initial conditions are specified in table
5.2.}\smallskip}}
\endinsert

\midinsert
\vbox{\centerline{%
\hskip4em\epsfxsize=.6\hsize\epsfbox{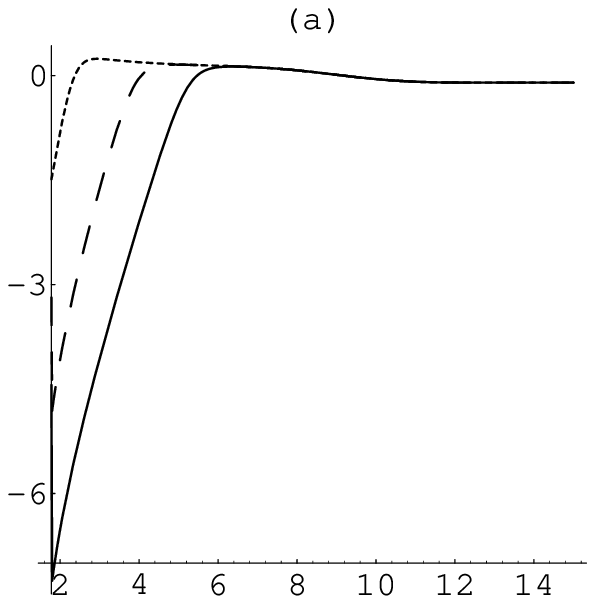}\hskip-10em\epsfxsize=.6\hsize\epsfbox{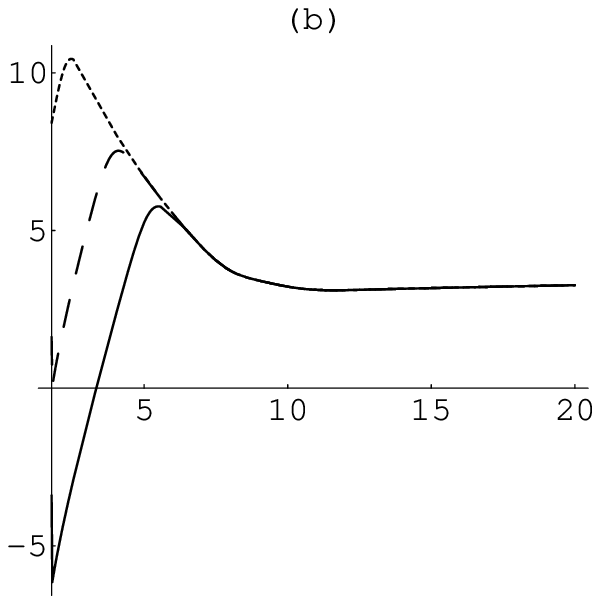}}
\vskip-.75in
{\baselineskip 10pt\noindent\narrower\rm\hbox{\eightbf
Figure 5.4}:\quad\eightrm
{Loop production
during the approach to scaling of an axionic string network. The horizontal
axis is labeled in orders of magnitude in time from the moment of string
formation, and plots represent the fraction of the energy in long strings
converted into loops per Hubble time (a) and the number of loops created per
Hubble volume per Hubble time (b). We have taken $\alpha\sim q$ (see
(\axcase)).}\smallskip}}
\endinsert

It should be noted that although string dynamics is friction-dominated until $t_{\star}$, the
same is not true for the corresponding term in the evolution equation for L (\evl) (this is
of course due to the $v^2$ dependence of that term). In fact the expansion term in (\evl) is
always dominant, although in the Kibble regime the friction and loop production terms
are of the same order of magnitude. Also note that, as we said before, the loop formation term
is proportionally more important in the Kibble regime  than in the `free' scaling regime---in
fact, the (linear-log) plot of the ratio of the friction and expansion terms in (\evl) is
indistinguishable (apart from the scale of the y axis) from figure 5.1(a)!

Axionic and GUT strings reach the final scaling regime well before $t_{eq}$, so we will separately
consider the approach to scaling in each case and then the transition between the linear scaling
regimes in the radiation and matter eras.

Figures 5.3-4 depict the approach to scaling of an axionic string network. The
discussion is analogous to the one for the early evolutionary stages of
electroweak strings, except for the different orders of magnitude involved.
The other difference is of course that the additional logarithmic dependencies
affect all the transient scaling laws with the exception of $L\propto
t^{1/2}$. Note that in all cases the effect of the logarithms is to increase
the exponent of the power-law dependencies. The stretching regime can now last
up to five orders of magnitude in time.

\subbegin{C. Global and gauge GUT strings}
\bigskip

The approach to scaling for a global GUT string network is shown in figure
5.5; as we mentioned previously, in this case loop reconnections onto the
network are not important and hence the cases plotted are representative of all
physically meaningful initial conditions.
In this case the period of friction-dominated evolution is much shorter, and so the stretching and
Kibble regimes, although still clearly distinguishable, are not very definite. Again, the
effect of the logarithmic corrections for global strings is to increase the power-law
dependencies. It takes a global GUT string network about six orders of magnitude in time to reach
the scaling regime.

Now, let us concentrate on the more interesting case of gauge GUT strings for which the approach
to the linear regime is shown in figures 5.6-7. Here the epoch when friction
dominates the dynamics lasts less than three orders of magnitude in time, and
the linear regime is reached about six orders of magnitude after string
formation. Hence the linear scaling regime is approached faster than
previously estimated and, more importantly, strings become relativistic at
a very early epoch. Consequently, energy losses to loops are relevant at all
times (see figure 5.7). One should also notice the similarity between the
plots of the exponent of the power-law dependence of the lengthscale L
(labeled (a) in 5.6) and of the fraction of the energy density in the form of
long strings converted into loops per Hubble time (labeled (a) in 5.7). This
is evidence of the fact that loop formation is the crucial mechanism for
cosmological string network evolution. It should be emphasized that with our
ansatz for $\alpha$,
$\alpha_{sc}\sim10^{-3}$ and $\Gamma G\mu\sim 65\times10^{-6}$ it takes about
4 orders of magnitude in time for $\alpha$ to evolve from $\alpha=1$ to
$\alpha=\alpha_{sc}$, and during this time the physical size of the loops
formed changes by less than a factor of 3. This is in agreement with both
simulations\refto{bb} and previous analytical estimates\refto{aca}.

Note that for the given parameters the energy density in string loops
in the linear regime is in fact larger than that in the long strings. In this case, our
analytical estimate was $\varrho_{est}\sim2.1$ in the radiation era, while numerically we find
$\varrho_{num}\approx3.26$ (see figure 5.7(c)). Furthermore, while with our
simplifying assumptions one finds that in the radiation-era scaling regime loops
decay at a time
$t_{decay}\sim5.18t_f$ after formation (see figure 5.7(d)), the exact result is
$t_{decay}\approx6.22t_f$. This is explained by the fact that in the analytic
estimate we have assumed that loops always have $v_\ell=1/sqrt{2}$, while
numerically we assume that their initial velocity is that of the long string
network at the moment of formation. These discrepancies get smaller as we decrease
the parameter $\alpha$ (hence the loop size at formation).

\midinsert
\vbox{\centerline{%
\hskip4em\epsfxsize=.6\hsize\epsfbox{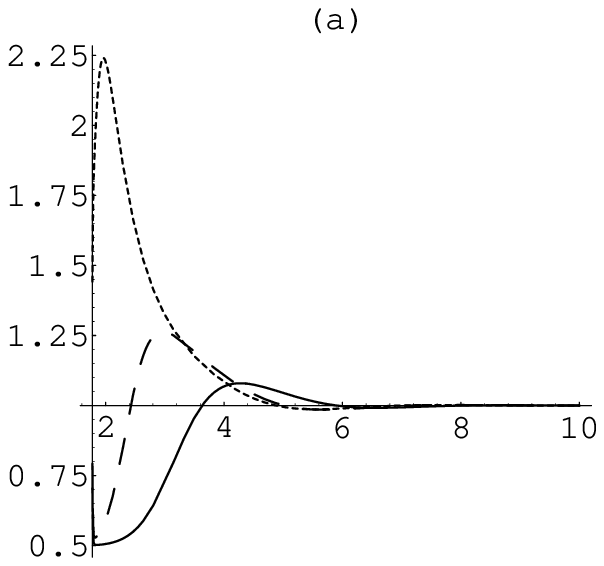}\hskip-10em\epsfxsize=.6\hsize\epsfbox{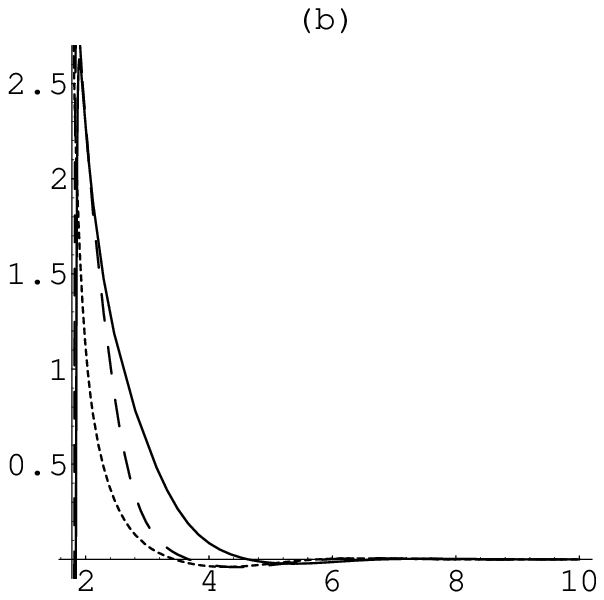}}
\vskip-.75in}
\endinsert
\midinsert
\vbox{\centerline{%
\hskip4em\epsfxsize=.6\hsize\epsfbox{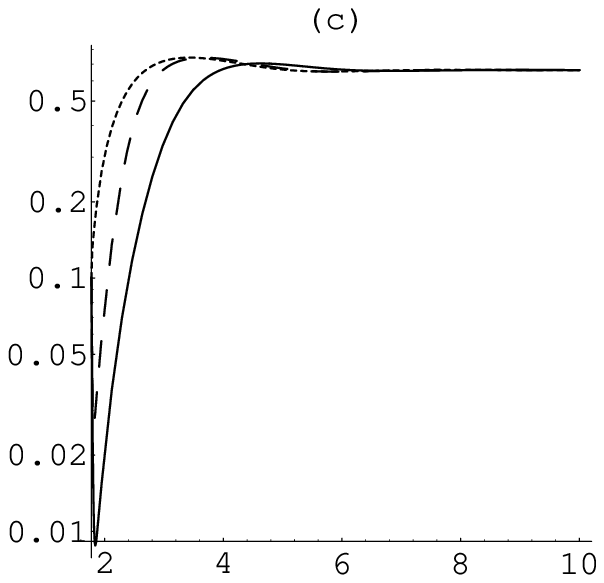}\hskip-10em\epsfxsize=.6\hsize\epsfbox{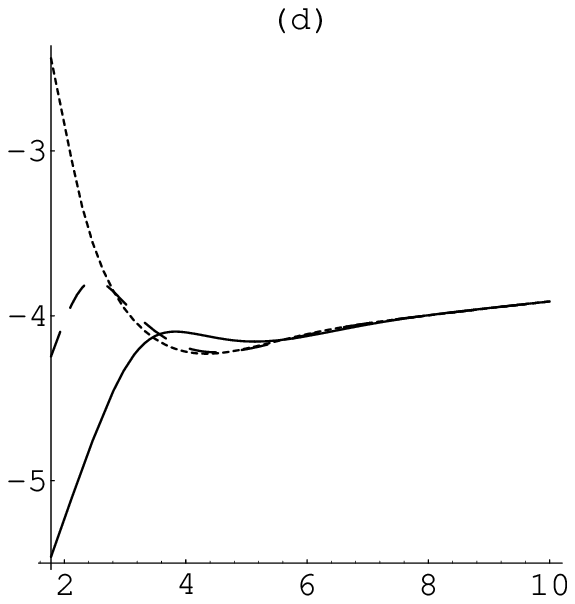}}
\vskip-.75in
{\baselineskip 10pt\noindent\narrower\rm\hbox{\eightbf
Figure 5.5}:\quad\eightrm
{The approach to
scaling of a global GUT string network. Plots successively represent the
exponents of the power-law dependence of $L$ (a) and $v$ (b), $v$ itself (c),
and the ratio of the long string and background densities (d). The horizontal
axis is labeled in orders of magnitude in time from the moment of string
formation, and initial conditions are specified in table 5.2.}\smallskip}}
\endinsert

It should also be
pointed out that although less loops are produced in the friction-dominated
regime, they live much longer---up to four orders of magnitude in time for
loops produced just after the network forms. In fact, with our choice of
parameters (note that $\alpha$ is close to the numerical upper limit) no
loop decays during the friction-dominated epoch, so the energy density in
loops relative to that in long strings grows to a value much larger that
the final value in the linear regime. Hence if we had included the loops
present at the moment of the network formation (which account for about 20\%
of the string length) we would find that loop density is always a
non-negligible fraction of the long-string density.  As one would expect, the loop population
takes a longer time to reach the scaling regime compared to the long-string network. 

Finally, we discuss the transition between the radiation and matter era linear scaling regimes (see
figures 5.8-10). Using our ansatzes for $k$ and $\alpha$, and assuming a smooth
change of ${\tilde c}$ (although this is strictly not necessary given the
relatively weak dependence of the scaling properties on ${\tilde c}$) we can match
all results the from numerical simulations\refto{bb}  and our own analytical
estimates (see section 4c) with an error of less than 10$\%$. This is a significant
achievement given the fact that in previous analytic models the discrepancies were
typically of order 100$\%$.

\midinsert
\vbox{\centerline{%
\hskip4em\epsfxsize=.6\hsize\epsfbox{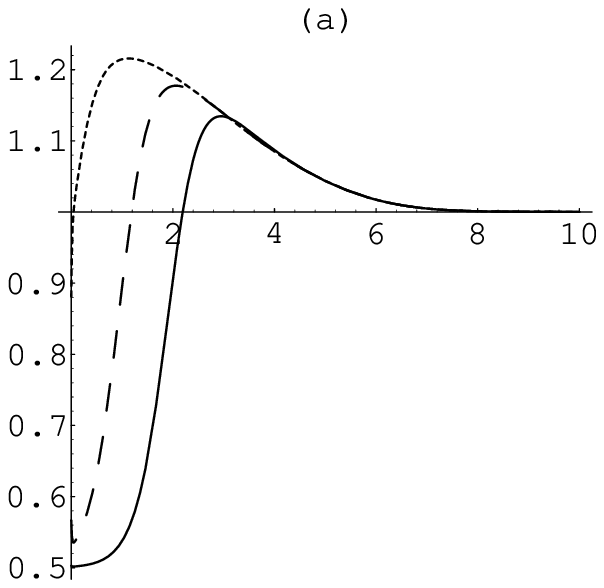}\hskip-10em\epsfxsize=.6\hsize\epsfbox{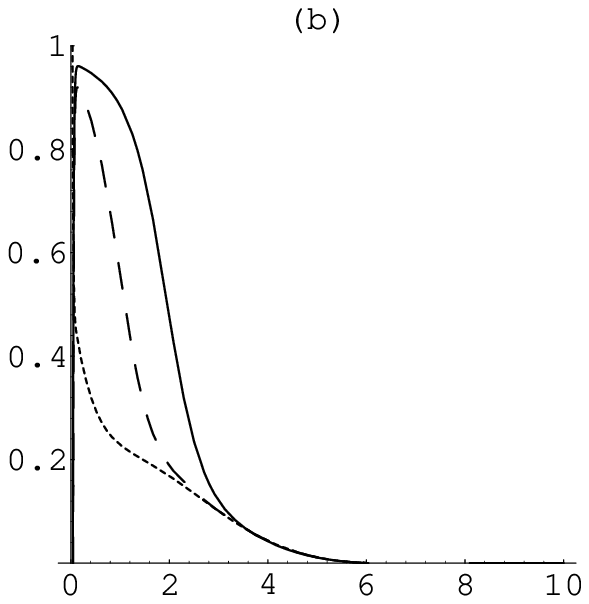}}
\vskip-.75in}
\endinsert
\midinsert
\vbox{\centerline{%
\hskip4em\epsfxsize=.6\hsize\epsfbox{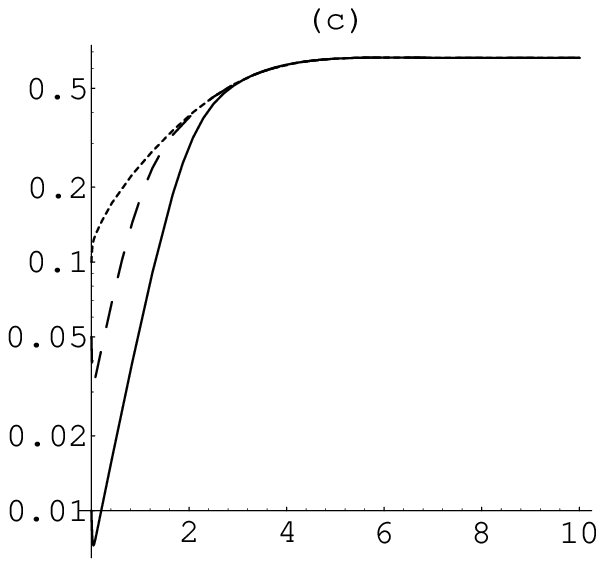}\hskip-10em\epsfxsize=.6\hsize\epsfbox{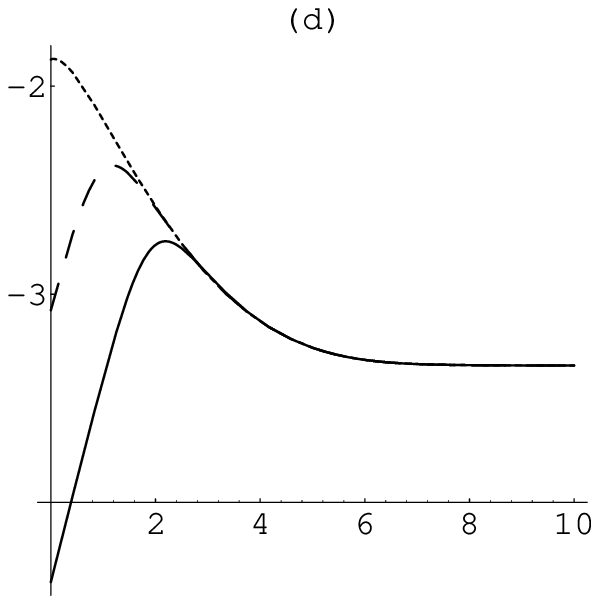}}
\vskip-.75in
{\baselineskip 10pt\noindent\narrower\rm\hbox{\eightbf
Figure 5.6}:\quad\eightrm
{The approach to
scaling of a gauge GUT string network. Plots successively represent the
exponents of the power-law dependence of $L$ (a) and $v$ (b), $v$ itself (c),
and the ratio of the long string and background densities (d). The horizontal
axis is labeled in orders of magnitude in time from the moment of string
formation, and initial conditions are specified in table 5.2.}\smallskip}}
\endinsert

Note that this is a much slower and smoother process than previously 
estimated, extending for about eight orders of magnitude in time. This is the
reason why Bennett
\& Bouchet\refto{bb}, having studied this transition with a series of numerical
simulations covering a range of five orders of magnitude in time, do not see the
scaling parameters "going flat". In fact, one can say that the string network never
leaves the scaling regime---the exponent of the power-law dependence of
$L$ is never more than ten percent away from the scaling value. It is also
interesting to observe that the string velocity increases just after $t_{eq}$,
before it decreases to the matter-era scaling value (see fig. 5.8(d)); this is due
to the comparatively rapid change in the expansion rate.

\midinsert
\vbox{\centerline{%
\hskip4em\epsfxsize=.6\hsize\epsfbox{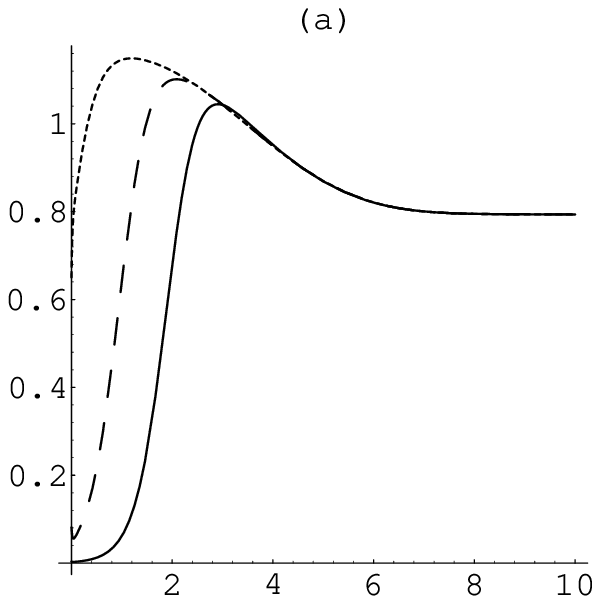}\hskip-10em\epsfxsize=.6\hsize\epsfbox{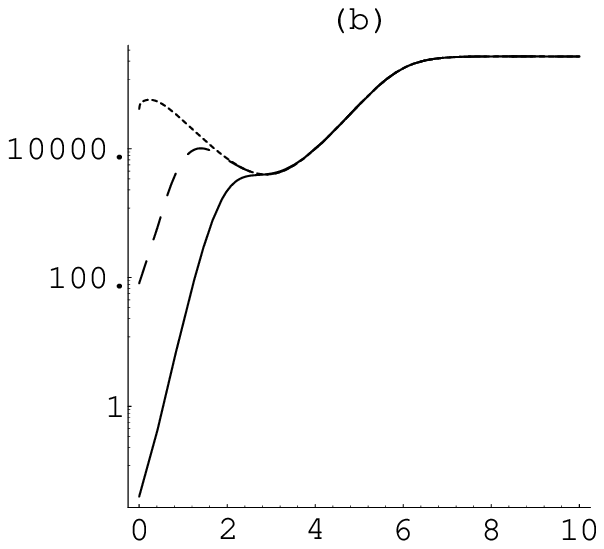}}
\vskip-.75in}
\endinsert
\midinsert
\vbox{\centerline{%
\hskip4em\epsfxsize=.6\hsize\epsfbox{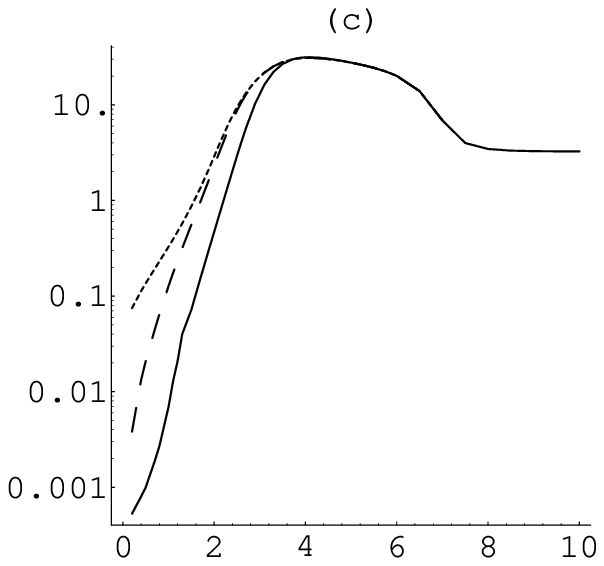}\hskip-10em\epsfxsize=.6\hsize\epsfbox{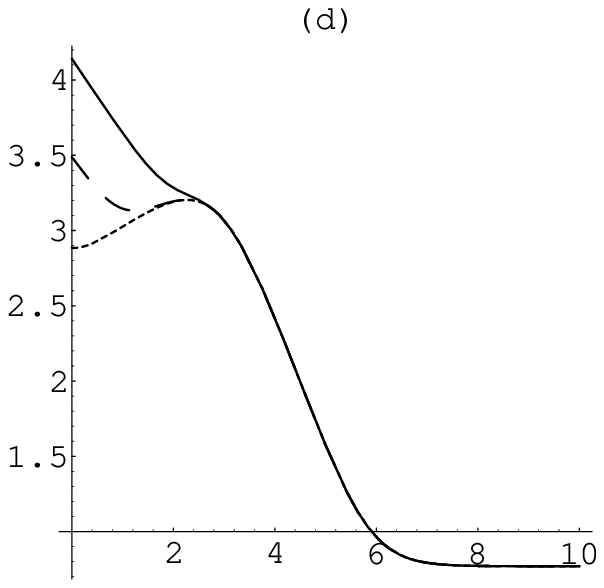}}
\vskip-.75in
{\baselineskip 10pt\noindent\narrower\rm\hbox{\eightbf
Figure 5.7}:\quad\eightrm
{Evolution of the
loop population for a gauge GUT string network. Plots successively represent
the fraction of the energy in long strings converted into loops per Hubble
time (a), the number of loops created per Hubble volume per Hubble time (b),
the ratio of loop and long string densities (c) and the time of loop decay (in
orders of magnitude after formation) for loops produced at each time (d). Note
that (a,b,d) were obtained from our analytical estimates (see section 4C)
while (c) is the (exact) numerical solution of (\ra2tio). The horizontal axis
is labeled in orders of magnitude in time from the moment of string
formation, and we have taken
$\alpha\sim10^{-3}$ and
$\Gamma G\mu\sim65\times10^{-6}$.}\smallskip}}
\endinsert

Note the effect of the logarithmic corrections for global strings in the evolution of the ratio of
the long string and background energy densities (see figure 5.9). It should also be
said that due to its specific parameter dependencies, the number of loops
produced  per Hubble volume per Hubble time is the quantity for which our
temporary matching problem is more serious.

To conclude, it should be emphasized that this quantitative picture of GUT string evolution
can lead to important modifications in the structure formation scenarios involving cosmic
strings.

\midinsert
\vbox{\centerline{%
\hskip4em\epsfxsize=.6\hsize\epsfbox{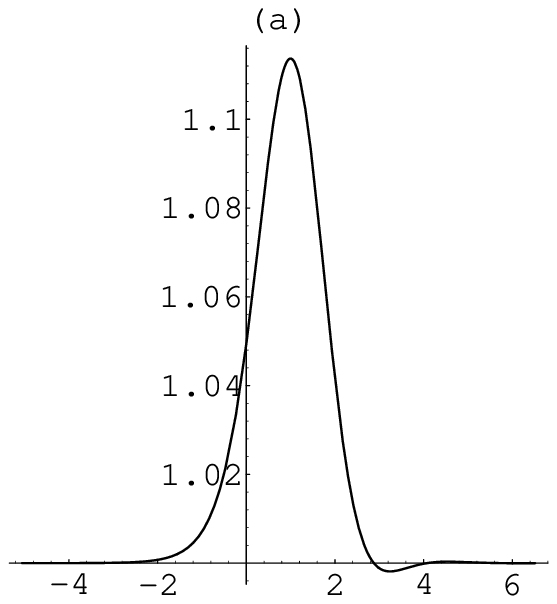}\hskip-10em\epsfxsize=.6\hsize\epsfbox{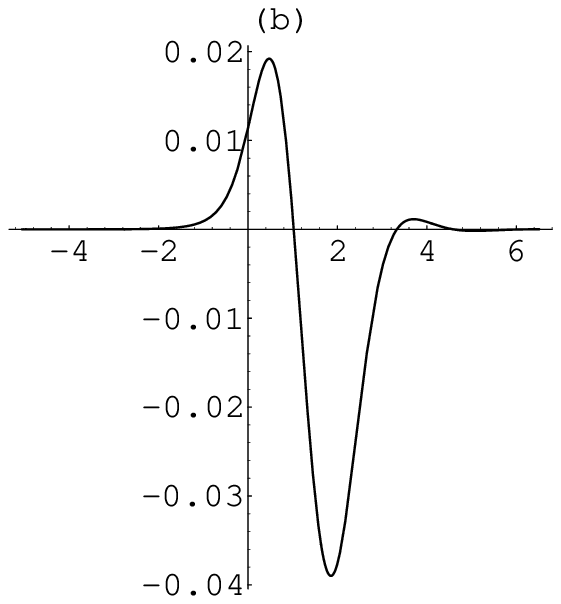}}
\vskip-.75in}
\endinsert
\midinsert
\vbox{\centerline{%
\hskip4em\epsfxsize=.6\hsize\epsfbox{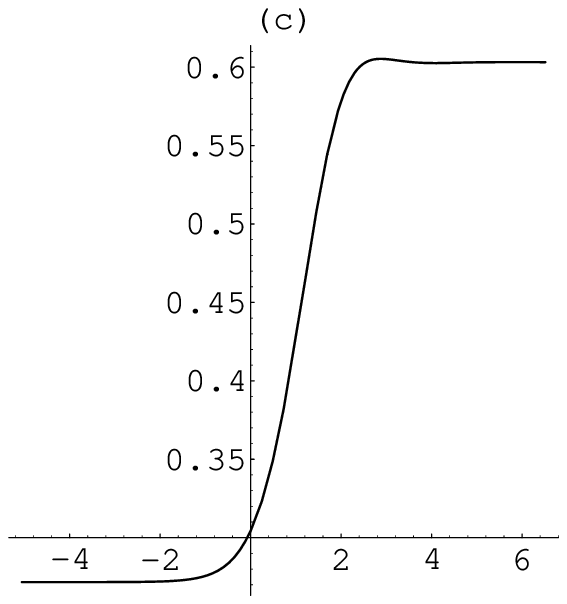}\hskip-10em\epsfxsize=.6\hsize\epsfbox{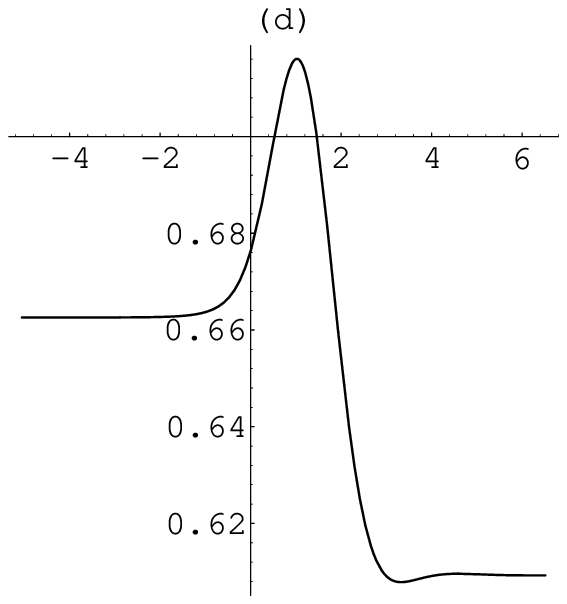}}
\vskip-.75in
{\baselineskip 10pt\noindent\narrower\rm\hbox{\eightbf
Figure 5.8}:\quad\eightrm
{Evolution of GUT
long-string networks (both gauge and global) in the transition between the radiation and matter
linear scaling regimes. Plots respectively represent the exponent of the power-law dependence
of $L$ (a) and $v$ (b), the ratio $L/t$ (c) and velocity (d). The horizontal axis is labeled in
terms of the logarithm of the scale factor (with $a(t_{eq})=1$); it spans the period between
$10^{-10}t_{eq}$ and $10^{10}t_{eq}$.}\smallskip}}
\endinsert

\midinsert
\vbox{\centerline{%
\hskip4em\epsfxsize=.6\hsize\epsfbox{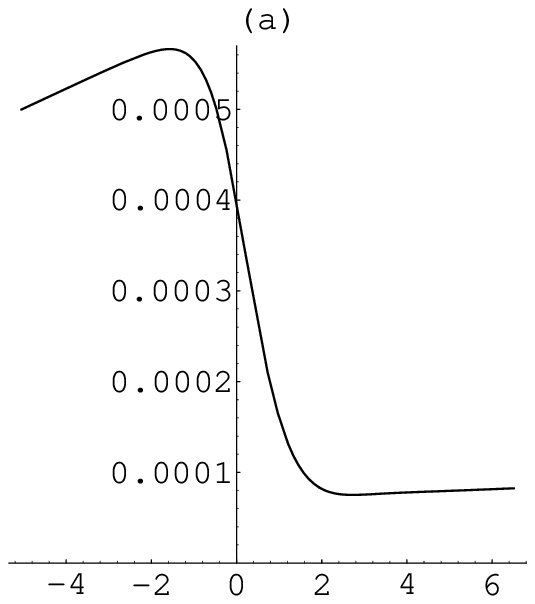}\hskip-10em\epsfxsize=.6\hsize\epsfbox{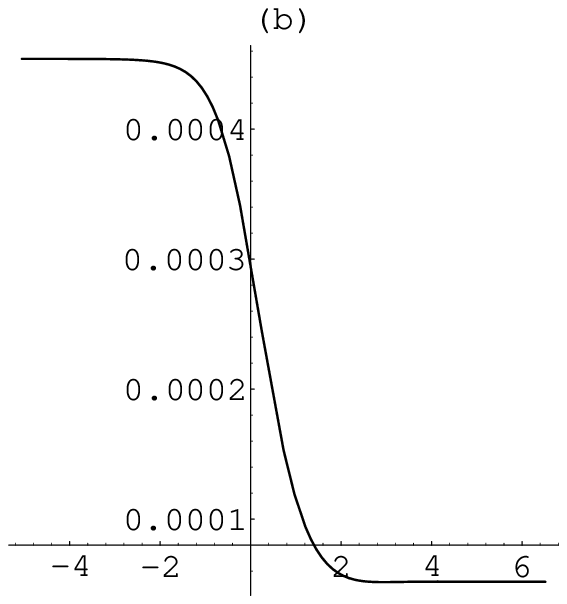}}
\vskip-.75in
{\baselineskip 10pt\noindent\narrower\rm\hbox{\eightbf
Figure 5.9}:\quad\eightrm
{Evolution of
the ratio of long string and background densities for global (a) and gauge (b)
GUT string networks in the transition between the radiation and matter eras.
The horizontal axis is labeled in terms of the logarithm of the scale factor
(with $a(t_{eq})=1$); it spans the period between
$10^{-10}t_{eq}$ and $10^{10}t_{eq}$.}\smallskip}}
\endinsert

\midinsert
\vbox{\centerline{%
\hskip4em\epsfxsize=.6\hsize\epsfbox{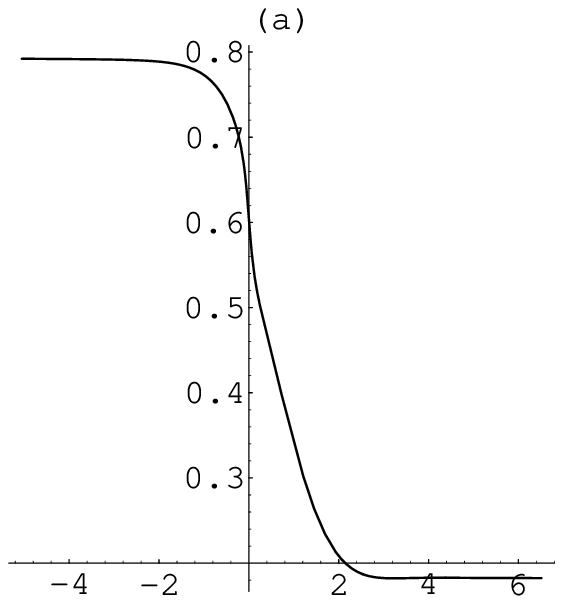}\hskip-10em\epsfxsize=.6\hsize\epsfbox{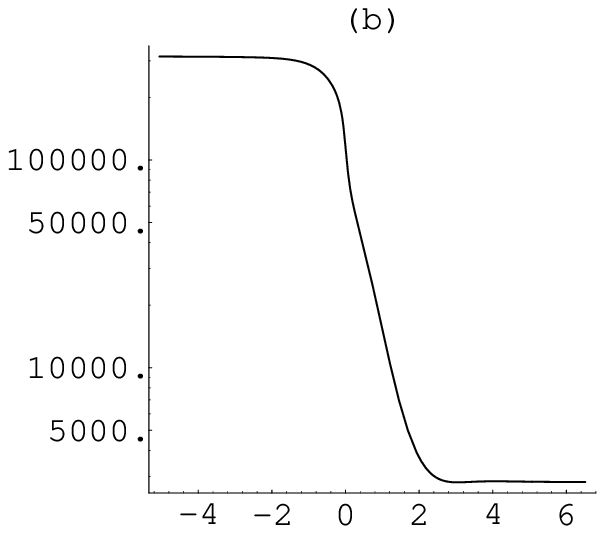}}
\vskip-.75in
{\baselineskip 10pt\noindent\narrower\rm\hbox{\eightbf
Figure 5.10}:\quad\eightrm
{Evolution of
loop characteristics for GUT string networks in the transition between the
radiation and matter eras. Plots represent the fraction of the energy density
in long string converted into loops per Hubble time (a) and the number of
loops produced per Hubble volume per Hubble time (b). The horizontal axis is
labeled in terms of the logarithm of the scale factor (with $a(t_{eq})=1$);
it spans the period between
$10^{-10}t_{eq}$ and $10^{10}t_{eq}$; we have taken
$\alpha\sim10^{-3}$ and $\Gamma G\mu\sim65\times10^{-6}$.}\smallskip}}
\endinsert

\sectbegin{6}{Conclusions}
\bigskip
\nobreak

In this paper we have presented a detailed account of a recently-developed
generalized `one-scale' model of string network evolution\refto{ms,ms2} where
a `characteristic lengthscale (the `correlation length' or average
inter-string distance in the case of long strings, the physical loop length in
the case of loops) and the average velocity are the dynamical variables.

The immediate benefit of this generalization is that one is thus able
to properly describe string motion in  friction-dominated contexts. As a consequence, this
simple model is the first complete and fully quantitative study of the evolution of a string
network and the corresponding loop population in both condensed matter
(see\refto{ms2}) and cosmological contexts.

Notably, we established the validity of Kibble's scaling law for intermediate
and light energy scale cosmic strings; furthermore, if the initial string density is
sufficiently low, there can also be an earlier period during which the strings are conformally
stretched. However, despite the relative growth in the string energy
$\rho_\infty/\rho_b$ in the damped epoch, we showed that strings can never dominate the energy
density of the universe. The model also predicts that electroweak strings are only approaching
the linear scaling regime today, whereas GUT strings reach this scaling regime faster than
previously  estimated. Finally, we have shown that the transition between the radiation and
matter era linear scaling regimes is a very slow process.

Application of the model to string loops has also led to the determination
of the evolution of the density and other relevant properties of the loop
population. We can use this model to
determine the loop density and other relevant properties at all times. We
have found that for typical values of the parameters characterizing loop
production and decay there is more
energy density in loops than in long strings.

These results can significantly affect some cosmological string scenarios.
The insight gained on the friction-dominated epoch of string evolution also calls for a
re-examination of the vorton problem\refto{vor} and of string-induced
baryogenesis\refto{barios}. It should also be possible to obtain more accurate estimates of the
contribution of axions to dark matte, and, of course, our results on the radiation-matter
transition for GUT strings can affect structure formation scenarios. We hope to tackle some of
these problems in the very near future.

We have also pointed out that the study of the evolution of cosmic string networks has been
an effort involving both analytic and numerical work. Our results concerning the
friction-dominated epoch of string evolution make it clear that a numerical study of this epoch
is long overdue, not least because (as was pointed out by Vilenkin\refto{v1}) including the frictional force is simplicity itself: it amounts to a
redefinition of the Hubble parameter,  as can be seen
from (\spc,\tc). There is also a need for a more careful study of the loop
population. We have made several predictions regarding the scaling ratios of the loop and long
string densities as a function of parameters characterizing the loop production and decay;
these should be testable by (slightly modified) existing numerical simulations.

Apart from the effect of reconnections back onto the long string network, which is
non-negligible in some cases, the outstanding remaining problem is that of the small-scale
structure seen in the simulations\refto{bb}. We believe that the parameter $k$, defined in
(\dfk), will ultimately provide a phenomenological means of introducing small-scale structure
effects in this model. Some steps in that direction have already been taken in this essay---in
particular, a possible solution for the famous `matching' problem of Kibble's one-scale model
has already become apparent. In particular, it seems likely that we can relate $k$ to the
effective (`renormalized') string energy per unit length
$\tilde\mu$, which is numerically seen to provide a good description of small-scale structure.
These important issues are presently being considered.

\nosectbegin{Acknowledgments}

\noindent We are grateful for the hospitality of the Isaac Newton Institute where some of these
problems were raised during the {\it Topological Defects} workshop, July--December, 1994. 
C.M.\ is funded by JNICT (Portugal) under `Programa PRAXIS XXI' (grant no. PRAXIS
XXI/BD/3321/94).  E.P.S.\ is funded by PPARC and we both acknowledge the support of PPARC and the
EPSRC, in particular the Cambridge Relativity rolling grant (GR/H71550) and a Computational Science
Initiative grant (GR/H67652).


\def\hang{}



\def\jnl#1#2#3#4#5#6{\hang{#1 [#2], {\it #4\/} {\bf #5}, #6.\par}
									}


\def\jnlerr#1#2#3#4#5#6#7#8{\hang{#1 [#2], {\it #4\/} {\bf #5}, #6;
{Erratum:} {\it #4\/} {\bf #7}, #8.\par}
									}


\def\prep#1#2#3#4{\hang{#1 [#2], `#3', #4.\par}
									}


\def\book#1#2#3#4{\hang{#1 [#2], {\it #3\/} (#4).\par}
									}

\def\genref#1#2#3{\hang{#1 [#2], #3.\par}
									}


\def\prl{Phys.\ Rev.\ Lett.}
\def\pr{Phys.\ Rev.}
\def\pl{Phys.\ Lett.}
\def\np{Nucl.\ Phys.}

\def\rmp{Rev.\ Mod.\ Phys.}
\def\cmp{Comm.\ Math.\ Phys.}

\def\jetp{Sov.\ Phys.\ JETP}

\nosectbegin{References}

\references

\baselineskip 10pt
\let\it=\nineit
\let\rm=\ninerm
\let\bf=\ninebf
\rm

\refis{ms}
\jnl{Martins, C.J.A.P., \& Shellard, E.P.S.}{1996}{}{\pr}{D53}{575 (R1)}

\refis{ms2}
\prep{Martins, C.J.A.P., \& Shellard, E.P.S.}{1995}{Averaged methods for vortex-string
evolution}{in preparation}

\refis{ack}
\jnl{Austin, D., Copeland, E.J., \& Kibble, T.W.B.}{1993}{Evolution of cosmic
string configurations}{\pr}{D48}{5594} and references therein.

\refis{v1}
\jnl{Vilenkin, A.}{1991}{Cosmic string dynamics with friction}{\pr}{D43}{1060}

\refis{VV}
\jnl{Vachaspati, T., \& Vilenkin, A.}{1984}{Formation and evolution of
cosmic strings}{\pr}{D30}{2036}

\refis{tb}
\jnl{Turok, N., \& Bhattacharjee, P.}{1989}{}{\rmp}{61}{1}

\refis{dsmag}
\jnl{Davis, R.L., \& Shellard, E.P.S.}{1989}{Global strings and superfluid
vortices}{\prl}{63}{2021}

\refis{vor}
\jnl{Davis, R.L. \& Shellard, E.P.S.}{1989}{Cosmic vortons}{\np}{B249}{557}

\refis{cdty}
\jnl{Chuang, I. {\it et al.}}{1991}{Cosmology in the laboratory: Defect
dynamics in liquid crystals}{Science}{251}{1336}

\refis{gs}
\jnl{Garriga, J., \& Sakellariadou, M.}{1993}{Effects of friction on
cosmic strings}{\pr}{D48}{2502}

\refis{aca}
\jnl{Allen, B., \& Caldwell, R.R.}{1991}{Kinky structure on strings}{\pr}{D43}{2457}

\refis{aus}
\jnl{Austin, D.}{1993}{Small-scale structure on cosmic
strings}{\pr}{D48}{3422}

\refis{k85}
\jnlerr{Kibble, T.W.B.}{1985}{Evolution of a system of cosmic
strings}{\np}{B252}{227}{B261}{750}

\refis{kd}
\jnl{Kibble, T.W.B.}{1986}{String-dominated universe}{\pr}{D33}{328}

\refis{b1}
\jnlerr{Bennett, D.P.}{1986}{Evolution of cosmic strings}{\pr}{D33}{872}{D34}{3932}
\jnl{Bennett, D.P.}{1986}{Evolution of cosmic strings. II}{\pr}{D34}{3592}

\refis{hk}
\jnl{Hindmarsh, M.B., \& Kibble, T.W.B.}{1995}{}{Rep. Prog. Phys.}{58}{477}

\refis{bb}
\jnl{Bennett, D.P., \& Bouchet, F.R.}{1990}{High-resolution simulations of
cosmic-string evolution. I. Network evolution}{\pr}{D41}{2408}
\jnl{Allen, B., \& Shellard, E.P.S.}{1990}{Cosmic-string evolution: A numerical
simulation}{\prl}{64}{119}

\refis{rsa}
\genref{Rohm, R.}{1985}{Ph.D. thesis, Princeton University.}
\jnl{de Sousa Gerbert, P. \& Jackiw, R.}{1988}{Classical and quantum scattering on a
spinning cone}{\cmp}{124}{229} \jnl{Alford, M.G. \& Wilczek, F.}{1989}{Aharonov-Bohm
interaction of cosmic strings with matter}{\prl}{62}{1071}

\refis{epa}
\jnl{Everett, A.E.}{1981}{Cosmic strings in unified gauge theories}{\pr}{D24}{858}
\jnl{Perkins, W.B. {\it et al.}}{1991}{Scattering of fermions from a cosmic string}{\np}{B353}{237}

\refis{albrecht}
\jnl{Albrecht, A. \& Turok, N.}{1989}{}{\pr}{D40}{973}

\refis{barios}
\jnl{Brandenberger, R.H. , Davis, A.-C. \& Trodden, M.}{1994}{Cosmic strings and
electroweak baryogenesis}{\pl}{B335}{123}

\refis{metc}
\jnl{Mermin, D.}{1979}{}{\rmp}{51}{591}

\refis{lctw}
\book{de Gennes, P.}{1981}{The physics of liquid crystals}{Clarendon Press, Oxford}
\jnl{Bowick, M. {\it et al.}}{1994}{}{Science}{263}{947}

\refis{cdty}
\jnl{Chuang, I. {\it et al.}}{1991}{Cosmology in the laboratory: Defect
dynamics in liquid crystals}{Science}{251}{1336}

\refis{zurek}
\jnl{Zurek, W.H.}{1985}{}{Nature}{317}{505}
\jnl{Hendry, P.C. {\it et al.}}{1994}{}{Nature}{368}{315}

\refis{msgv}
\jnl{Salomaa, M. \& Volovik, G.}{1987}{}{\rmp}{59}{533}
\jnl{Parts, U. {\it et al.}}{1995}{}{\prl}{75}{3320}

\refis{abrk}
\jnl{Abrikosov, A.}{1957}{}{\jetp}{5}{1174}

\endreferences

\end